\documentstyle[12pt,epsf]{article}
\begin{document}
\title{Monopole-Skyrmions}

\author{{\large B. Kleihaus}$^{\dagger}$,
{\large D. H. Tchrakian}$^{\dagger \star}$ 
and {\large F. Zimmerschied}$^{\dagger}$ \\ \\
$^{\dagger}${\small Department of
Mathematical Physics, National University of Ireland Maynooth,} \\
{\small Maynooth, Ireland} \\ \\
$^{\star}${\small School of Theoretical Physics -- DIAS, 10 Burlington Road,
Dublin 4, Ireland }}

\date{}
\newcommand{\dd}{\mbox{d}}\newcommand{\tr}{\mbox{tr}}
\newcommand{\ee}{\end{equation}}
\newcommand{\be}{\begin{equation}}
\newcommand{\ii}{\mbox{i}}\newcommand{\e}{\mbox{e}}
\newcommand{\pa}{\partial}\newcommand{\Om}{\Omega}
\newcommand{\vep}{\varepsilon}
\newcommand{\bfph}{{\bf \varphi}}
\newcommand{\lm}{\lambda}
\newcommand{\Pa}{\Psi_a}
\newcommand{\Ph}{\Psi_h}
\newcommand{\Pf}{\Psi_f}
\newcommand{\sff}{\sin\hspace*{-0.5ex} f}
\newcommand{\cff}{\cos\hspace*{-0.5ex} f}
\newcommand{\sqff}{\sin^2\hspace*{-0.5ex} f}
\newcommand{\cqff}{\cos^2\hspace*{-0.5ex} f}
\newcommand{\sqqff}{\sin^4\hspace*{-0.5ex} f}
\newcommand{\cqqff}{\cos^4\hspace*{-0.5ex} f}
\newcommand{\hk}{\kappa^2 }
\def\theequation{\arabic{equation}}
\renewcommand{\thefootnote}{\fnsymbol{footnote}}
\newcommand{\re}[1]{(\ref{#1})}
\newcommand{\bfR}{{\sf R\hspace*{-0.9ex}\rule{0.15ex}%
{1.5ex}\hspace*{0.9ex}}}
\newcommand{\N}{{\sf N\hspace*{-1.0ex}\rule{0.15ex}%
{1.3ex}\hspace*{1.0ex}}}
\newcommand{\Q}{{\sf Q\hspace*{-1.1ex}\rule{0.15ex}%
{1.5ex}\hspace*{1.1ex}}}
\newcommand{\C}{{\sf C\hspace*{-0.9ex}\rule{0.15ex}%
{1.3ex}\hspace*{0.9ex}}}
\newcommand{\1}{{\sf 1\hspace*{-0.75ex}\rule{0.03ex}%
{1.4ex}\hspace*{0.75ex}}}
\renewcommand{\thefootnote}{\arabic{footnote}}
\renewcommand{\textfraction}{0.9}
\renewcommand{\arraystretch}{1.5}

\maketitle
\begin{abstract}
A systematic numerical study of the classical solutions to the
combined system consisting of the Georgi-Glashow model and the $SO(3)$
gauged Skyrme model is presented. The gauging of the Skyrme system permits
a lower bound on the energy, so that the solutions of the composite system
can be topologically stable. The solutions feature some very interesting
bifurcation patterns, and it is found that some branches of these
solutions are unstable.
\end{abstract}
\medskip
\medskip
\newpage

\section{Introduction}

The physical motivation of the present work is to set the framework for
a semiclassical approach to describing the mechanism of monopole
catalysis of Baryon number decay, proposed by Rubakov~\cite{R2} and
Callan~\cite{C}. Here we are led by the work of Callan and
Witten~\cite{CW}, where the Baryon is described as the soliton of the
Skyrme~\cite{S} model, in the backgroud of the $U(1)$ Maxwell field of 
a monopole in the Dirac gauge. 
In the present work, the Baryon is again described by a
Skyrmion, which in this case interacts with the full $SO(3)$ non-Abelian
Higgs model (the Georgi-Glashow model) so that the Skyrmion we consider
is gauged with the full $SO(3)$ group and interacts with the
't~Hooft-Polyakov monopole~\cite{tp}. We will refer to this model as 
Monopole-Skyrmion model (MSM).

The most important difference between the present work and that of
Ref.~\cite{CW} is that here we have {\it two} distinct topological charges
- the first being the Baryon number of the Skyrmion and the second, the
monopole charge. The energy of our composite system therefore has a
topological lower bound consisting of a combination of these two charges
of rather different geometric natures, the Baryon charge being a
{\it degree} which cannot be expressed as a total divergence while the
monople charge is a {\it flux} by virtue of being descended from the
second Chern-Pontryagin class.

The most important feature of describing the interaction of the Skyrmion
with the gauge field in our Monopole-Skyrmion model is 
the presrciption of gauging the Skyrme field with the diagonal $SU(2)$. 
This prescription was introduced in Ref.~\cite{AT} for the gauged Skyrme model,
where no Higgs field and Higgs potential are present,  
and the resulting solutions were studied in Ref.~\cite{BT,BKT}.
Most importantly, this gauging permits a lower bound on the energy of 
the gauged Skyrmion unlike in the case when the usual gauging is from the left, 
e.g. in Ref.~\cite{DF} where there is no lower bound.
In this paper, we shall refer to the models arising from the
gauging prescription used in \cite{AT} as gauged Skyrme models (GSM).
Since the topological lower bounds for the GSMs were presented in detail
in Refs.~\cite{AT,BT}, and because all that we need to know here is that
these exist, we do not discuss them further here.

Now the presence of a topological lower bound is not a sufficient
condition for the existence of a topologically stable soliton.
To illustrate this we refer to the graph of the energy versus Skyrme coupling 
constant in Fig.~1 for the GSM studied in~\cite{BT,BKT}. Without being 
mathematically rigorous one can suppose that
the branches $A_{\rm gS}$ and $B_{\rm gS}$ correspond to solutions, which form
local minima of the energy functional, whereas the connecting 
branch $\tilde A_{\rm gS}$ correspond to solutions, which form sattlepoints.

\begin{figure}[!h]
\centering
\mbox{\epsfysize=12cm\epsffile{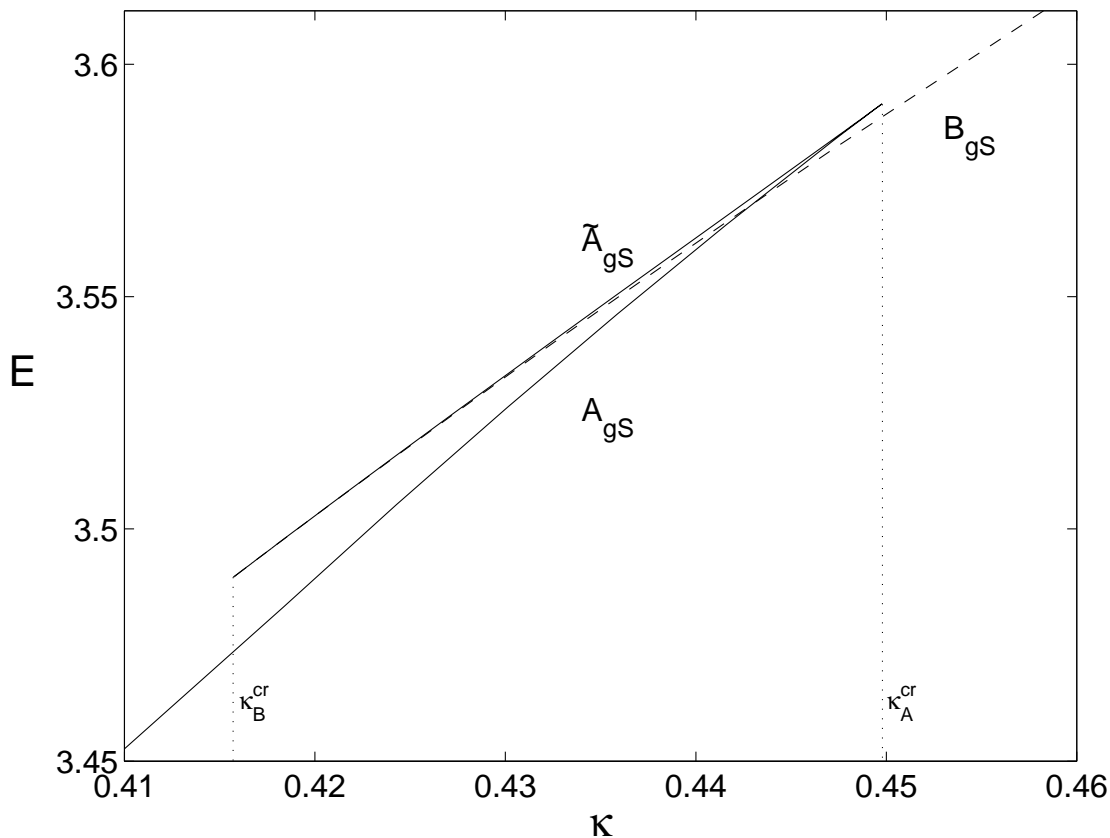}}
\caption{ 
The energy of the gauged Skyrmion as a function of $\kappa$ in the
region of the bifurcation.
}
\end{figure}

The bifurcation structure in Fig.~1 is quite different from that
appearing in models where the Skyrme field is gauged as in \cite{DF}
without a topological lower bound. The corresponding graph to Fig. 1 in
that case is given in the work of Ref.~\cite{EKS} featuring only two
branches as opposed to the three in Fig. 1. Of these two branches
\cite{EKS} only one corresponds to stable solutions, as expected from the
work of Refs.~\cite{R1}. Thus a butterfly pattern of bifurcations with two
stable branches seems to be typical of of GSMs with topological lower bounds.
We will find in our study of the Monopole-Skyrme model, that the
butterfly of Fig. 1 persits for some range of the parameters in the model.

Having already presented the lower bound on the energy of Monopole-Skyrme model 
in Ref.~\cite{BKT}, we do not repeat it here and proceed straight away to 
the study the solutions numerically, with a view to exposing some of their 
qualitative features that may be of some physical interest. 
The model we employ here is a slightly modified version of the one in 
Ref.~\cite{BKT}. It features a particular interaction term
between the Higgs field and the chiral field in such a way that it also
fixes the asymptotic values of the chiral field subject to the finite
energy condition, and assuring integral Baryon number. In Section 2 we
present the model, discuss its spectrum, and discuss some some scaling
properties which will become pivotal in the subsequent numerical analysis.
In Section 3, we subject the system to spherical symmetry and give the
classical equations to be integrated, the detailed numerical results of
which we give in Section 4. In Section 5 we analyse the normal modes of
the radial fluctuations around the solutions to the MSM, characterised
by those values of the parameters for which the solutions display butterfly
bifurcations, and verify that indeed the branches corresponding to
$\tilde A_{\rm gS}$ are unstable. A summary and
discussions of our results are given in Section 6.

\section{The Model}

The Lagrangian of the Monopole-Skyrme model is given by

\begin{eqnarray}
-{\cal L}[A,\Phi,U] & = & 
\int\left(
 \frac{1}{2g^2} Tr\left\{ F_{\mu\nu} F^{\mu\nu}\right\}
+\frac{1}{4} Tr\left\{D_\mu \Phi D^\mu \Phi\right\}
+\frac{\tilde{\lambda}}{2} Tr\left\{(\Phi^2-\eta^2)^2\right\}
\right.
\nonumber
\\
& &
\left.
-\frac{f_\pi^2}{4} Tr\left\{ D_\mu U U^\dagger D^\mu U U^\dagger\right\}
-\frac{\tilde{\kappa}^2}{8} 
Tr\left\{ \left[D_\mu U U^\dagger, D_\nu U U^\dagger\right]
          \left[D^\mu U U^\dagger, D^\nu U U^\dagger\right]\right\}
\right.
\nonumber
\\
& &
\left.
+V(\Phi, U) \right) d^3r
\label{lag1}
\end{eqnarray}
with 
\begin{equation}
V(\Phi, U) = \frac{\tilde{g}_{\pi\Phi}^2f_\pi^2}{4} 
Tr\left\{ 
 \Phi^2-\left[\frac{1}{4}\left(\{\Phi,U\}+\{\Phi,U^\dagger\}\right)\right]^2
\right\} \ .
\label{Vpot}
\end{equation}
The field strength tensor of the $su(2)$ gauge potential $A_\mu = A_\mu^a
\frac{\tau_a}{2}$
is defined as
\begin{eqnarray}
F_{\mu\nu} & = & \partial_\mu A_\nu - \partial_\nu A_\mu 
                  + i \left[A_\mu, A_\nu \right] \ ,
\label{Fdef}
\end{eqnarray}
and the covariant derivatives 
for the Higgs field $\Phi = \phi^a \tau_a$ and the 
chiral matrix $U=\exp\{i\pi^a \tau_a\}$ are defined as
\begin{eqnarray}
D_\mu \Phi & = & \partial_\mu \Phi + i \left[ A_\mu, \Phi \right] \ , 
\label{DPdef}
\\
D_\mu U & = & \partial_\mu U + i \left[ A_\mu, U \right] \ , 
\label{DUdef}
\end{eqnarray}
respectively.
$g$ denotes the gauge coupling parameter, 
$\lambda$ the strength of the Higgs potential,
$\eta$ the norm of the vacuum expectation value of the Higgs field,
$f_\pi$ the pion decay constant and
$\kappa$ the Skyrme coupling parameter.  The parameter
$g_{\pi\Phi}$ characterises the direct coupling between the Higgs boson and 
the chiral matrix.

The first three terms in  (\ref{lag1}) are the familiar Georgi-Glashow
model which is characterised by the scale $\eta$. 
The next two terms are the Skyrme model with covariant derivatives 
$D_\mu U$ allowing
for an interaction of the chiral matrix with the gauge potential. 
The first of these terms introduces another scale, $f_\pi$.
The last term $V(\Phi,U)$ describes a direct coupling of the chiral matrix with
the Higgs field, on which we will comment later.

The Lagrangian (\ref{lag1}) is invariant under local $SU(2)$ 
gauge transformations ${\bf g}$, 
\begin{eqnarray}
A_\mu &\longrightarrow & {\bf g} A_\mu {\bf g}^{-1} 
+ i \partial_\mu {\bf g} {\bf g}^{-1} \ ,
\nonumber\\
\Phi  &\longrightarrow & {\bf g} \Phi {\bf g}^{-1} \ ,
\nonumber\\
U  &\longrightarrow & {\bf g} U  {\bf g}^{-1}\ .
\nonumber 
\end{eqnarray} 

The vacuum of the theory is given by the following constant configuration,
\begin{equation}
A_\mu \equiv 0\ , \ \Phi \equiv \eta \tau_3 \ , \ U \equiv \1\ .
\label{vac}
\end{equation}
In order to identify the particle content of the model
we expand around the vacuum,
\begin{equation}
A_\mu = \delta a_\mu^a \frac{\tau_a}{2}\ , \ 
\Phi = \eta \tau_3 +\delta \phi^a\tau_a\ , 
\ U = (\1- \frac{\delta\pi^a\delta\pi^a}{2}) + i \delta\pi^a\tau_a \ ,
\label{expvac}
\end{equation}
and insert into the equation of motion obtained from the Lagrangian
(\ref{lag1}). Neglecting quadratic terms in $\delta a_\mu^a$, 
$\delta \phi^a$ and $\delta\pi^a$ we find
\begin{eqnarray}
\partial_\mu\partial^\mu \delta \pi^a & = & \tilde{g}_\pi^2\eta^2 \delta \pi^a \ , \ a=1,2,3 \ ,
\nonumber \\
\partial_\mu\partial^\mu \delta \phi^3 & = & 
4\tilde{\lambda}\eta^2 \delta \phi^3\ ,
\nonumber \\
\partial_\mu\partial^\mu \delta a^{\nu,3} & = & 0\ ,
\nonumber \\
\partial_\mu\partial^\mu \delta \bar{a}^{\nu,a} & = & g^2 \eta^2 \delta \bar{a}^{\nu,a}
\ , \ a=1,2 \ \ ,
\nonumber
\end{eqnarray}
where we have defined 
${\displaystyle 
\delta \bar{a}_\mu^a = \delta a_\mu^a 
+\frac{1}{\eta} \partial_\mu \delta \phi^b \epsilon_{a3b}}$,
and used the gauge fixing conditions $\partial_\mu \delta a^{\mu, 3}=0$ and   
$\partial_\mu \delta \bar{a}^{\mu, a}=0$ for $a=1,2$.
Thus we find for the masses of the gauge field, the chiral field and the 
Higgs field,
\begin{equation}
m_A = g \eta \ ,\ \ m_\pi = \tilde{g}_{\pi\Phi} \eta \ ,
 \ \ m_h = 2\sqrt{\tilde{\lambda}} \eta \ ,
\label{masses}
\end{equation}
respectively.
(We take all parameters $g,\tilde{g}_{\pi\Phi},\eta,\tilde{\lambda}$ 
to be positive).
Clearly, the mass of the chiral field stems form the interaction 
term of the chiral matrix with the Higgs field, which breaks the 
chiral symmetry. However, the spontaneous symmetry breaking does not 
lead to a mass splitting for the chiral fields.

In the following we will motivate the potential term (\ref{Vpot}) for the 
Higgs field and the chiral matrix.
First consider the model without the potential term.
>From the finite energy conditions we find for the chiral field and the 
Higgs field in the limit $r \rightarrow \infty$
\begin{equation}
D_\mu \Phi \rightarrow 0 \ , \ \ \ D_\mu U \rightarrow 0 \ . 
\label{E_infty}
\end{equation}
Let us assume, that this condition is fulfilled for the radial component
of the covariant derivatives and concentrate on the angular components.
We decompose the Higgs field and the chiral matrix in the 
form
\begin{eqnarray}
\Phi  & = & \ \ \ \ \ \ \ \  h \hat{\phi} \ , \ \ \ \ \ \ \ \   
\ \ \hat{\phi} = \hat{\phi}^a \tau_a  \ , \ \ {\rm with}
\ \ \hat{\phi^a} \hat{\phi^a} =1 \ ,
\nonumber\\
U_\infty & = & \cff\1 + i \sff \hat{u} \ , 
\ \ \hat{u} = \hat{u}^a \tau_a  \ , \ \ {\rm with}
\ \ \hat{u^a} \hat{u^a} =1 \ ,
\label{PU_dec}
\end{eqnarray}
respectively, where $h, \hat{\phi}^a$ and $f, \hat{u}^a$ are functions of 
the variables $r, \theta,\varphi$. 

>From the finite energy conditions follows that at infinity 
$|h|=\eta, f_\infty=const$. However, $\hat{\phi}^a$ and $\hat{u}^a$
may still be functions of the angular variables $\theta,\varphi$.
The conditions (\ref{E_infty}) now become 
\begin{equation}
D_\alpha \Phi_\infty
= \left. (\partial_\alpha \hat{\phi} 
+ i [ A_\alpha,\hat{\phi}])\right|_\infty = 0 \ , \ \
D_\alpha U_\infty
= \sff_\infty \left. \left (\partial_\alpha \hat{u} 
+ i [ A_\alpha,\hat{u}] \right)\right|_\infty
= 0 \ ,
\end{equation}
where $\alpha= \theta, \varphi$.
The first condition can be fulfilled with
$\left. A_\alpha \right|_\infty = 
\left. \left(\frac{i}{2} \partial_\alpha \hat{\phi}\hat{\phi} 
+ A_\alpha^{(\phi)} \hat{\phi}\right) \right|_\infty$, 
where $A_\alpha^{(\phi)}$ is 
some function of $\theta,\varphi$. The second condition can be fulfilled
either with $f_\infty=0$ or with
$\left. A_\alpha \right|_\infty  = 
\left. \left(\frac{i}{2} \partial_\alpha \hat{u}\hat{u} 
+ A_\alpha^{(u)} \hat{u}\right) \right|_\infty
$, where $A_\alpha^{(u)}$ is again some function of $\theta,\varphi$.
If the  latter condition is fulfilled, $f_\infty$ may take arbitrary values.
Hence, these configurations can be deformed continuously into configurations
with trivial chiral matrix.

In order to avoid this problem we have introduced the potential (\ref{Vpot})
into the Lagrangian.
Using the general decomposition (\ref{PU_dec}) the potential can be written as
$V(\Phi, U)
 = \frac{1}{2}\tilde{g}_{\pi\Phi}^2 f_\pi^2 h^2\sqff$, 
i.~e. it couples the modulus of the Higgs field $|h|$ 
and the chiral function $f$, and
does not depend on the ``phases'' $\hat{\phi}^a, \hat{u}^a$.
Consequently, the masses for the chiral fields, introduced by the potential
at infinity, do not depend on the direction of the Higgs field in isospace.
Furthermore, the finite energy condition now 
forces  $f_\infty = n \pi$, where $n$ is an integer. 

The sum of the potentials in (\ref{lag1}) has global minima 
at $|h|=\eta, f=n\pi$, and 
the matrix of the second variations has only non-negative eigenvalues at the 
global minima, provided the parameters $\lambda$ and $\tilde{g}_{\pi\Phi}^2$ 
are positive.

In order to study the consequences of the two scales $f_\pi$ and $\eta$
we will take two points of view.
First we will fix $f_\pi$ and express all dimensionful quantities in 
units of $f_\pi$. Equivalently, we can fix $\eta$ and express all dimensionful 
quantities in units of $\eta$. In each case, the ratio of the scales will
enter the equations of motion as a parameter.

We define the dimensionless quantities
$x = r f_\pi g\ , \ \ \hat{\Phi} = \Phi/f_\pi \ , 
\ \ \eta_0 = \eta/f_\pi \ $.
Then the Hamiltonian becomes 
\begin{eqnarray}
{\cal H}[A,\hat{\Phi},U] & = & \frac{f_\pi}{g}
\int\left(
 \frac{1}{2} Tr\left\{ F_{\mu\nu} F^{\mu\nu}\right\}
+\frac{1}{4} Tr\left\{D_\mu \hat{\Phi} D^\mu \hat{\Phi}\right\}
+\frac{{\lambda}}{2} Tr\left\{(\hat{\Phi}^2-\eta_0^2)^2\right\}
\right.
\nonumber
\\
& &
\left.
-\frac{1}{4} Tr\left\{ D_\mu U U^\dagger D^\mu U U^\dagger\right\}
-\frac{{\kappa}^2}{8} 
Tr\left\{ \left[D_\mu U U^\dagger, D_\nu U U^\dagger\right]
          \left[D^\mu U U^\dagger, D^\nu U U^\dagger\right]\right\}
\right.
\nonumber
\\ 
& &
\left.
+\frac{{g}_{\pi\Phi}^2}{4} 
Tr\left\{ 
 \hat{\Phi}^2-\left[\frac{1}{4}\left(\{\hat{\Phi},U\}
 +\{\hat{\Phi},U^\dagger\}\right)\right]^2
\right\}
\right) d^3{x} \ ,
\label{ham1}
\end{eqnarray}
where we have defined 
${\lambda} = \tilde{\lambda}/g^2\ , \ 
{\kappa} = \tilde{\kappa} g\ , \ 
{g}_{\pi\Phi} = \tilde{g}_{\pi\Phi}/g\ $.
In terms of masses Eq.~(\ref{masses}) the  parameters ${g}_{\pi\Phi}$ and 
$\lambda$ can be expressed as
${g}_{\pi\Phi}= m_\pi/m_A$ and $\sqrt{{\lambda}} = 2 m_h/m_A$, respectively. 
The parameter $\eta_0= f_\pi/\eta$ denotes the ratio of the scales.
Apart form the last term 
the Hamiltonian (\ref{ham1}) this is equivalent to the Hamiltonian 
studied before in \cite{BKT}. 
The difference is, that we now consider $\eta_0$ as a free parameter.
In the limit $\eta_0 \rightarrow 0$ the minimum of the Higgs potential 
allows for a vanishing Higgs field. 
In this case we find the gauged Skyrme model considered before in
\cite{AT,BT,BKT}.

Fixing the scale parameter $\eta$ we define
$\bar{x}=\eta g r$, $\bar{\Phi}= \eta \Phi$ and $\xi = f_\pi/\eta$.
Then the Hamiltonian becomes
\begin{eqnarray}
{\cal H}[A,\bar{\Phi},U] & = & \frac{\eta}{g}
\int\left(
 \frac{1}{2} Tr\left\{ F_{\mu\nu} F^{\mu\nu}\right\}
+\frac{1}{4} Tr\left\{D_\mu \bar{\Phi} D^\mu \bar{\Phi}\right\}
+\frac{\bar{\lambda}}{2} Tr\left\{(\bar{\Phi}^2-1)^2\right\}
\right.
\nonumber
\\
& &
\left.
-\frac{\xi^2}{4} Tr\left\{ D_\mu U U^\dagger D^\mu U U^\dagger\right\}
-\frac{\bar{\kappa}^2}{8} 
Tr\left\{ \left[D_\mu U U^\dagger, D_\nu U U^\dagger\right]
          \left[D^\mu U U^\dagger, D^\nu U U^\dagger\right]\right\}
\right.
\nonumber
\\ 
& &
\left.
+\frac{\bar{g}_{\pi\Phi}^2\xi^2}{4} 
Tr\left\{ 
 \bar{\Phi}^2-\left[\frac{1}{4}\left(\{\bar{\Phi},U\}
 +\{\bar{\Phi},U^\dagger\}\right)\right]^2
\right\}
\right) d^3\bar{x} \ ,
\label{ham2}
\end{eqnarray}
with $\bar{\lambda}=\tilde{\lambda}/g^2= {\lambda}$, 
$\bar{\kappa}=\tilde{\kappa} g={\kappa}$
and 
$\bar{g}_{\pi\Phi}=\tilde{g}_{\pi\Phi}/g= {g}_{\pi\Phi}$.

Comparing with the case of fixed scale $f_\pi$ we find 
$\xi=1/\eta_0$, $\bar{x}=\eta_0 {x}$, $\bar{\Phi}=\hat{\Phi}/\eta_0$.
Because the Hamiltonians (\ref{ham1}) and (\ref{ham2}) are equivalent, 
we can obtain the properties of (\ref{ham1}) from (\ref{ham2}) and vice versa 
by using these relations. In particular, for the dimensionless energies
${\displaystyle 
{E}\equiv \frac{g}{4\pi f_\pi}  {\cal H}}$
and
${\displaystyle 
\bar{E}\equiv \frac{g}{4\pi\eta }  {\cal H}}$
we have ${\displaystyle \bar{E}= \frac{E}{\eta_0}}$. 
We will opt to work with (\ref{ham1}) in the following.

\section{Static spherically symmetric equations}
The static spherically symmetric, purely magnetic Ansatz for 
the gauge field is \cite{RY,AKY1}
\begin{equation}
A_0=0 \ , 
A_r =c(x)\frac{\tau_r}{2}\ , 
\ \ A_\theta = \left(1-a(x)\right) \frac{\tau_\varphi}{2} 
+ b(x)\frac{\tau_\theta}{2} , 
\ \ A_\varphi = -\sin\theta\left(\left(1-a(x)\right)\frac{\tau_\theta}{2} -
                            b(x)\frac{\tau_\varphi}{2}\right)\  \ ,
\label{A_an}
\end{equation}
where the $su(2)$ matrices $\tau_\alpha$, $\alpha=r,\theta,\varphi$ are 
defined in terms of the Pauli matrices $\tau_1,\tau_2,\tau_3$ by
\begin{eqnarray}
\tau_r & = & \sin\theta(\cos\varphi \tau_1 + \sin\varphi \tau_2) 
             + \cos\theta \tau_3 \ , 
\nonumber\\             
\tau_\theta & = & \cos\theta(\cos\varphi \tau_1 + \sin\varphi \tau_2)
             - \sin\theta \tau_3 \ , 
\nonumber\\             
\tau_\varphi & = & -\sin\varphi \tau_1 + \cos\varphi \tau_2 \ .
\nonumber
\end{eqnarray}
The spherically symmetric Ansatz for the Higgs field is 
\begin{equation}
\hat{\Phi} = h(x) \tau_r \ ,
\label{Phi_an}
\end{equation}
and for the chiral matrix 
\begin{equation}
U = \cff(x) + i \sff (x) \tau_r \ .
\label{U_an}
\end{equation}
The Ansatz is form invariant under the $U(1)$ gauge transformations 
\cite{RY}
\begin{equation}
 {\bf g}= \exp{\left\{i\frac{\Gamma}{2} \tau_r\right\}} \ , 
\label{invarg}
\end{equation} 
where the gauge transformation function $\Gamma$ is an arbitrary function 
of $x$. 
The gauge field functions transform as
\begin{eqnarray}
c(x) & \rightarrow & c(x) - x \Gamma'(x) \ ,
\nonumber\\
b(x) & \rightarrow & \cos\Gamma(x) b(x) - \sin\Gamma(x) a(x)  \ ,
\nonumber\\
a(x) & \rightarrow & \cos\Gamma(x) a(x) + \sin\Gamma(x) b(x) \ ,
\label{invarf}
\end{eqnarray} 
whereas the Higgs field function $h(x)$ and the chiral function $f(x)$ are 
invariant. 
To fix this gauge freedom we first impose the condition $c(x)\equiv 0$,
which still allows for global transformations with $\Gamma = const$.
Further, we find that the functions $a(x)$ and $b(x)$ enter the 
Lagrangian only in the form $a'^2(x)+b'^2(x)$ and $a^2(x)+b^2(x)$,
which permits us to set  $b(x)\equiv 0$.

With the Ansatz (\ref{A_an})-(\ref{U_an}) restricted to the gauge fixing 
conditions and $b(x)\equiv 0$ the Hamiltonian (\ref{ham1}) becomes
\begin{eqnarray}
{\cal H}[a,h,f] & = & \frac{4 \pi f_\pi}{g} \int\left( 
a'^2+\frac{(a^2-1)^2}{2x^2} +\frac{x^2 h'^2}{2}+a^2h^2
+ {\lambda}(h^2-\eta_0^2)^2x^2
\right.
\nonumber\\
& &
\left.
+\frac{x^2f'^2}{2} + a^2 \sqff 
+ 4 {\kappa}^2 a^2 \sqff ( f'^2 +\frac{a^2 \sqff}{2x^2})
\right.
\nonumber\\
& &
\left.
+\frac{ {g}_{\pi\Phi}^2 x^2}{2}h^2 \sqff
\right) dx \ .
\label{H_sph}
\end{eqnarray}

The differential equations for the functions $a(x), h(x)$ and $f(x)$
can now be obtained as the variational equations which extremize
the Hamiltonian (\ref{H_sph}),
\begin{eqnarray}
a'' & = & a\left\{\frac{(a^2-1)}{x^2} + h^2 
+\sqff \left[ 1 + 4  {\kappa}^2 (f'^2+\frac{a^2\sqff}{x^2})\right]
\right\} \ , 
\nonumber\\
h'' & = & -2\frac{h'}{x} 
          +h \left\{ 2 \frac{a^2}{x^2} + 4{\lambda}(h^2-\eta_0^2)
          + {g}_{\pi\Phi}^2 \sqff\right\}\ , 
\nonumber\\
f''  & = & \left\{   
8  {\kappa}^2 a \sff\left[a \cff\left( \frac{a^2\sqff}{x^2}-f'^2\right)
                              -2 \sff a' f' \right]
+2(\sff \cff a^2 -x f')\right. 
\nonumber\\
& &
\left.
+ {g}_{\pi\Phi}^2 \sff \cff h^2 x^2\right\}
\frac{1}{x^2 +8 {\kappa}^2a^2\sqff}\ . 
\label{deqs0}
\end{eqnarray}

These equations have to be solved due to
boundary conditions which ensure regularity of the solution at 
the origin and finite energy, i.~e.
\begin{eqnarray}
x=0 & : & a=1 \ , \ \ h=0 \ , \ \ f=\pi \ , 
\nonumber \\
x \rightarrow \infty  & : & 
 a\rightarrow 0 \ , \ \  h \rightarrow \eta_0 \ , \ \ f \rightarrow 0 \ .
\label{BC}
\end{eqnarray}  

\section{Numerical results}

We have constructed numerically solutions of the model for several values
of the parameters $\eta_0$, $ {\lambda}$, $ {\kappa}$ and 
$ {g}_{\pi\Phi}$.
In particular we investigated the dependence of the solutions on the 
parameters  $\eta_0$ and $ {\kappa}$.

\begin{figure}[!h]
\centering
\mbox{\epsfysize=8.5cm\epsffile{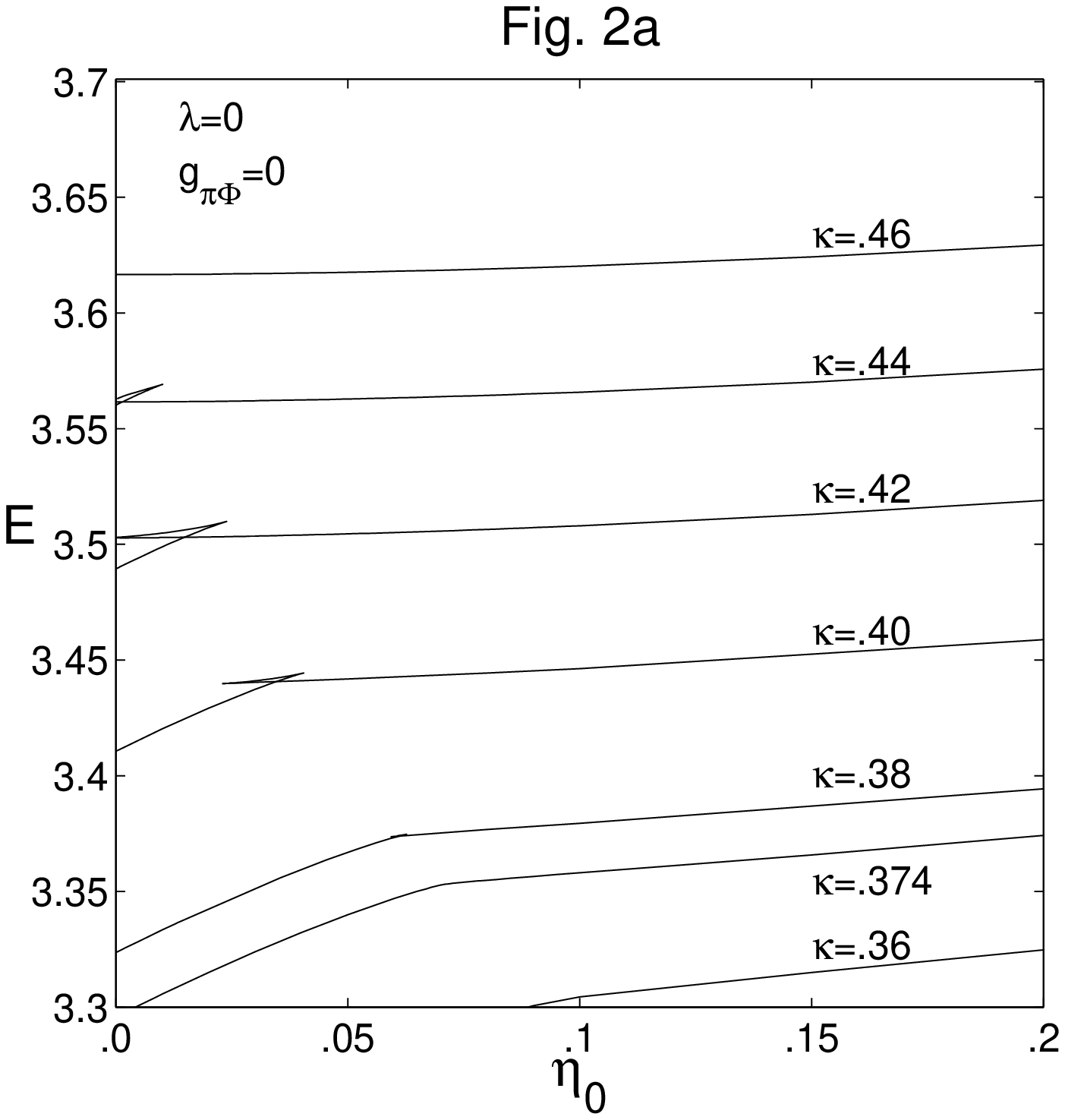}\epsfysize=8.5cm \epsffile{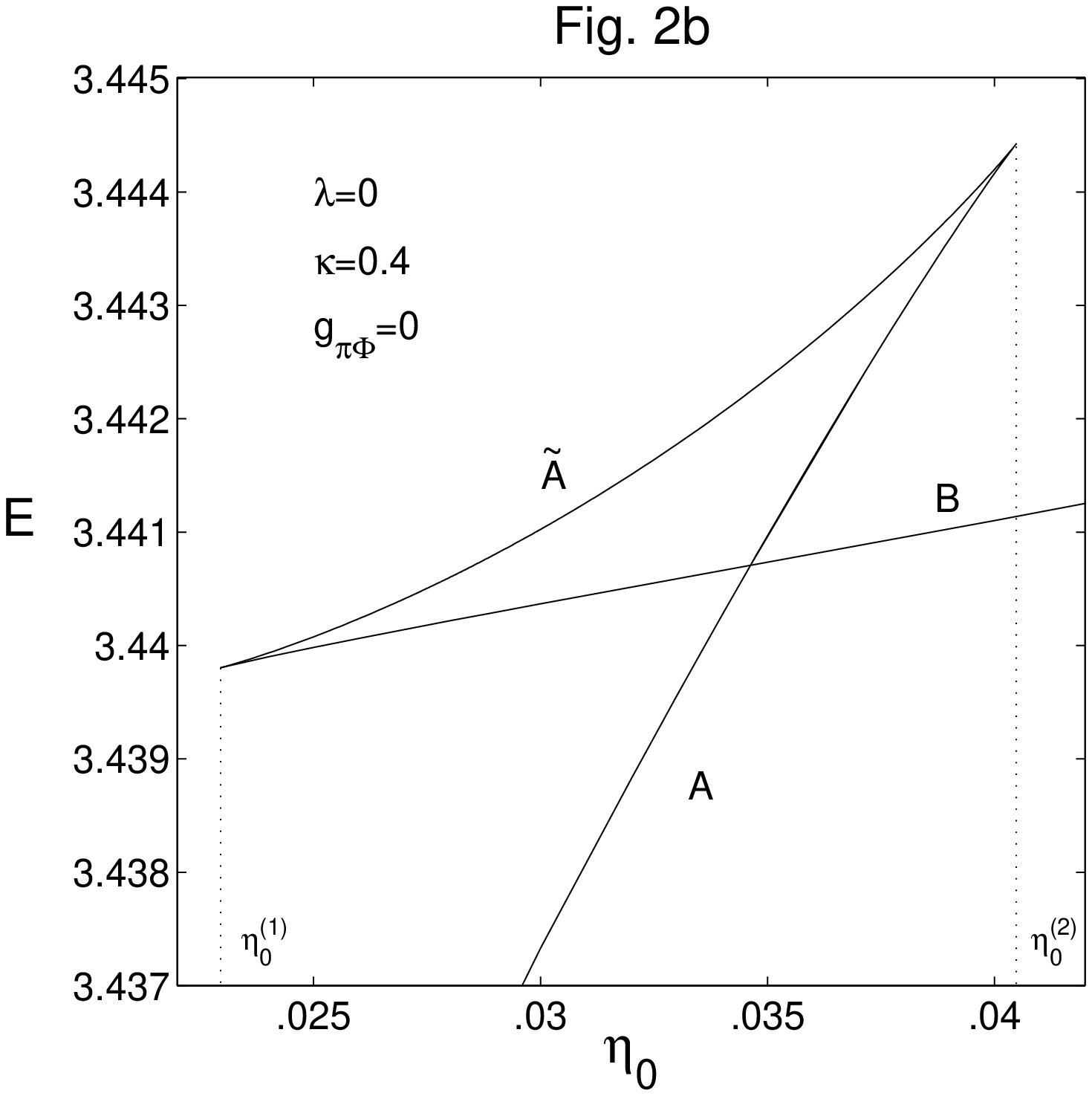}}
\caption{
(a) The dimensionless energy of the Monopole-Skyrmion solutions is shown 
as a function of $\eta_0$ for several values of $\kappa$ for 
fixed $\lambda=0$ and $g_{\pi\Phi}=0$.
(b)
The same as (a) for fixed $\kappa=0.4$, $\lambda=0$ and $g_{\pi\Phi}=0$.
$\eta_0^{(1)}$ and $\eta_0^{(2)}$  indicate the values of $\eta_0$ 
where the branches $B$ and $A$ merge with the saddlepoint branch $\tilde A$, 
respectively.}
\end{figure}

In Figs.~2a and 2b we show the energy 
${\displaystyle  {E} =\frac{g}{4\pi f_\pi} {\cal H}}$ as a function of
$\eta_0$ for several values of $ {\kappa}$ for $ {\lambda}=0$ and 
$ {g}_{\pi\Phi}=0$.
For large values of $\eta_0$ the energy is a monotonically increasing 
function of $\eta_0$. In the limit $\eta_0 \rightarrow \infty$ the energy
increases linearly with $\eta_0$, such that $ {E}/\eta_0\rightarrow 1$,
i.~e. the energy becomes equal to the Monopole energy
(in units of $4\pi \eta$). 
Indeed, this limit corresponds to the limit $f_\pi\rightarrow 0$ where
the chiral field becomes trivial, $U=-1$ everywhere except at infinity.

\subsection{$\eta \ll f_\pi$}

For small values of $\eta_0$ the solutions develop bifurcations, corresponding
to the `butterfly' structures in Fig.~2a, where for a fixed value of $\eta_0$
three solutions coexist.
We observe form Fig.~2a that the bifurcations occur only for a finite 
range of the parameter $ {\kappa}$, 
$ {\kappa}_{\rm cr}^{(1)} <  {\kappa}< {\kappa}_{\rm cr}^{(2)}$,
where $ {\kappa}_{\rm cr}^{(1)}, {\kappa}_{\rm cr}^{(2)}$ depend on
$ {\lambda}$ and $ {g}_{\pi\Phi}$. 
For $ {\lambda}=0$, $ {g}_{\pi\Phi}=0$
we find $ {\kappa}_{\rm cr}^{(1)}\approx 0.374$ and 
$ {\kappa}_{\rm cr}^{(2)}\approx 0.4495$. 
We demonstrate the details of the bifurcations in Fig.~2b for 
$ {\kappa}=0.4$ as an example.
This figure suggests that the branches $B$ and $A$ 
correspond to local minima of the energy functional, whereas the 
branch $\tilde{A}$ corresponds to saddlepoint solutions. 

The bifurcation pattern looks similar to the bifurcation pattern found 
recently in the gauged Skyrme model \cite{BKT}. 
Indeed, the bifurcations in the Monopole-Skyrme
model and the gauged Skyrme model are closely related to each other.
In the limit $\eta_0 \rightarrow 0$ the Higgs potential allows for a 
vanishing Higgs field. In this case we obtain the gauged Skyrmion model
studied in Refs.~\cite{AT,BT, BKT}. Consequently, the Monopole-Skyrmion
solutions
should approach the gauged Skyrmion solution in the limit of vanishing $\eta_0$.
In Fig.~3 we show in a 3D graph the energy of the Monopole-Skyrmion solutions
as a function of $\eta_0$ and $ {\kappa}$ together with the energy of the 
gauged Skyrmion solutions. We observe, that indeed 
the energy of the Monopole-Skyrmion coincides with the energy of the 
gauged Skyrmion in the limit $\eta_0 \rightarrow 0$.

\begin{figure}[!h]
\centering
\mbox{\epsfxsize=12.cm\epsffile{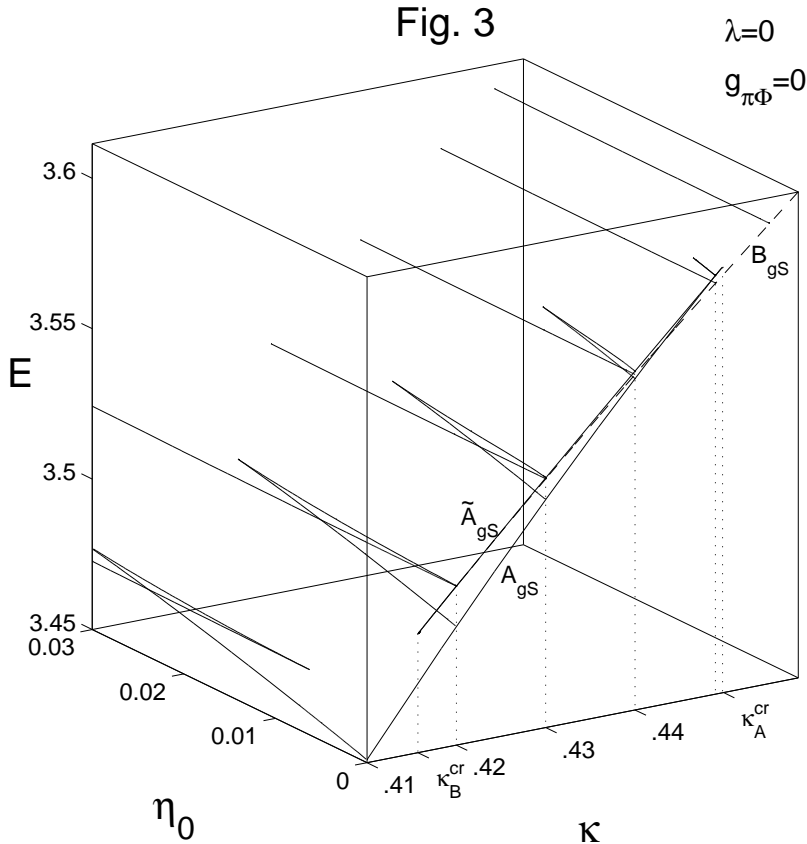}}
\caption{
The dimensionless energy of the Monopole-Skyrmion solutions is shown 
as a function of $\eta_0$ for several values of $\kappa$ for 
fixed $\lambda=0$ and $g_{\pi\Phi}=0$
together with the dimensionless energy of the gauged Skyrmion solutions 
as a function of $\kappa$.
$\kappa_{\rm B}^{\rm cr}$ and $\kappa_{\rm A}^{\rm cr}$  indicate the values 
of $\kappa$ 
where the branches $B_{\rm gS}$ and $A_{\rm gS}$ merge with the 
branch $\tilde{A}_{\rm gS}$, respectively.}
\end{figure}

In order to understand the behaviour of the solutions for small $\eta_0$,
let us first discuss the solutions of the Higgs-less gauged-Skyrme model.
In this model the absence of the Higgs field allows two possibilities for the value of
the gauge field function $a(x)$ at infinity. For large values of $ {\kappa}$
the function $a(x)$ vanishes at infinity 
(branch $B_{\rm gS}$) whereas for small values of 
$ {\kappa}$ the function $a(r)$ approaches the value one at infinity
(branches $A_{\rm gS}$ and $\tilde{A}_{\rm gS}$).
These two cases correspond to the dashed and solid lines,
respectively, plotted in the $E- {\kappa}$ plane in Fig.~3.
For a certain range of values 
$ {\kappa}_{\rm B}^{cr} < {\kappa} <  {\kappa}_{\rm A}^{cr} $ 
three branches of solutions
are present, reminding one of two local minima 
($B_{\rm gS}$ and $A_{\rm gS})$ and 
a saddlepoint ($\tilde{A}_{\rm gS}$) of the energy functional.

Now consider the limit $\eta_0 \rightarrow 0$ for the Monopole-Skyrmion
solutions.
If $ {\kappa} <  {\kappa}_{\rm B}^{cr}$ the Monopole-Skyrmion
solution will approach the unique gauged Skyrmion solution represented by the 
solid line in Fig.~3, below $  \kappa_B^{cr}$. However, if 
$ {\kappa}_{\rm B}^{cr} < {\kappa} <  {\kappa}_{\rm A}^{cr} $ 
there are three different gauged Skyrmion solutions available. 
In this case each Monopole-Skyrmion solutions of the branches 
$A$, $\tilde{A}$ and $B$  approaches the corresponding gauged Skyrmion 
solution on the branches $A_{\rm gS}$, $\tilde A_{\rm gS}$ and $B_{\rm gS}$,
respectively.
For $ {\kappa} >  {\kappa}_{\rm A}^{cr}$ the 
Monopole-Skyrmion solutions on the branches $A$ and $\tilde{A}$ 
Monopole-Skyrmion solutions cease to exist and the solutions on the remaining
branch $B$ tend uniquely to the gauged-Skyrmion ($B_{\rm gS}$)
solutions represented by the dashed line above $  \kappa_A^{cr} $in Fig.~3.

The limit $\eta_0 \rightarrow 0$ is non-uniform for
$ {\kappa} < \kappa_A^{cr}$ for the Monopole-Skyrmion solutions on the branches
$A$ and $\tilde A$. This is expected because the asymptotic values
of the gauge field function $a(x)$ for the Monopole-Skyrmion solutions and the 
gauged-Skyrmion solutions (branches $A_{\rm gS}$ and $\tilde{A}_{\rm gS}$)
are different. For the latter the function $a(x)$ approaches the 
value one at infinity, whereas for the former ones $a(x)$ vanishes at infinity.
To illustrate the limit $\eta_0 \rightarrow 0$ for small values of 
$ {\kappa}$ we exhibit in Figs.~3a-3c  as an example a sequence of 
field configurations of Monopole-Skyrmion solutions along the ``butterfly''
for $ {\kappa}=0.4$ in Figs.~2a and 2b. 
We follow the branch $B$ down to $\eta_0^{(1)}$ (see Fig.~2b), 
continue with increasing $\eta_0$ on the saddlepoint branch $\tilde{A}$ 
up to $\eta_0^{(2)}$ and finally we approach $\eta_0=0$ on the branch $A$.
 
\begin{figure}[!h]
\centering
\mbox{\epsfxsize=8.5cm\epsffile{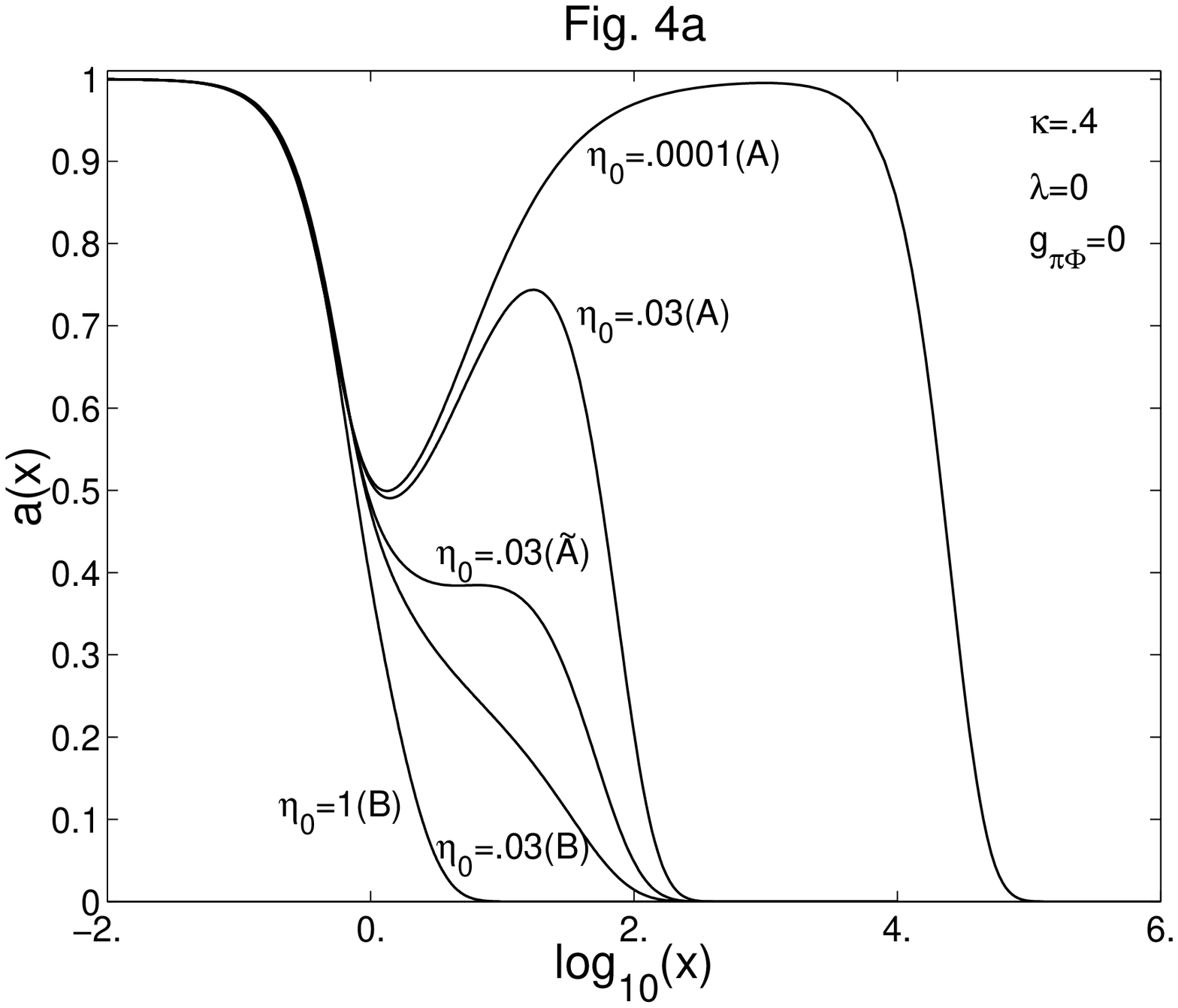}
      \epsfxsize=8.5cm\epsffile{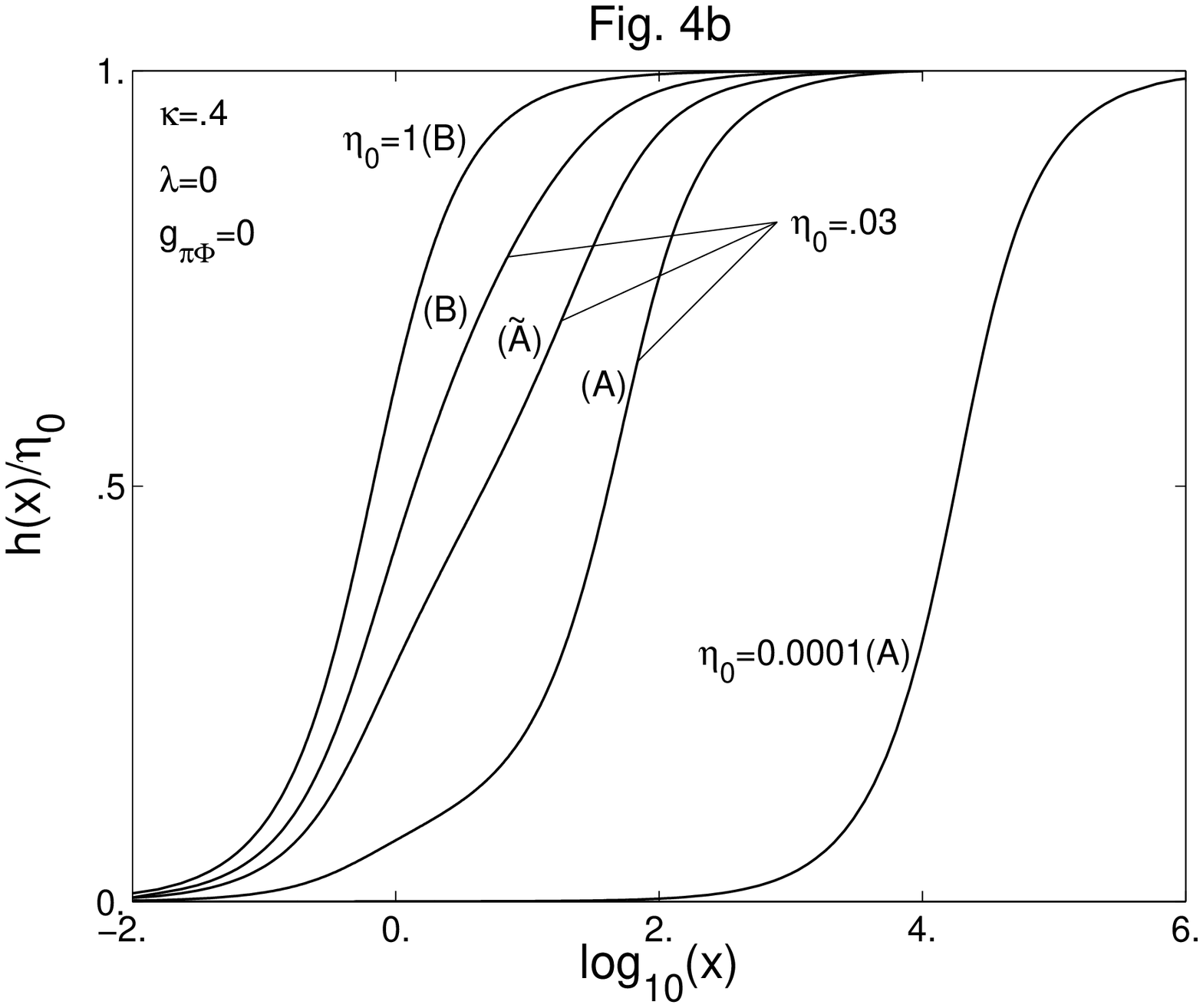}}\\
\mbox{\epsfxsize=8.5cm\epsffile{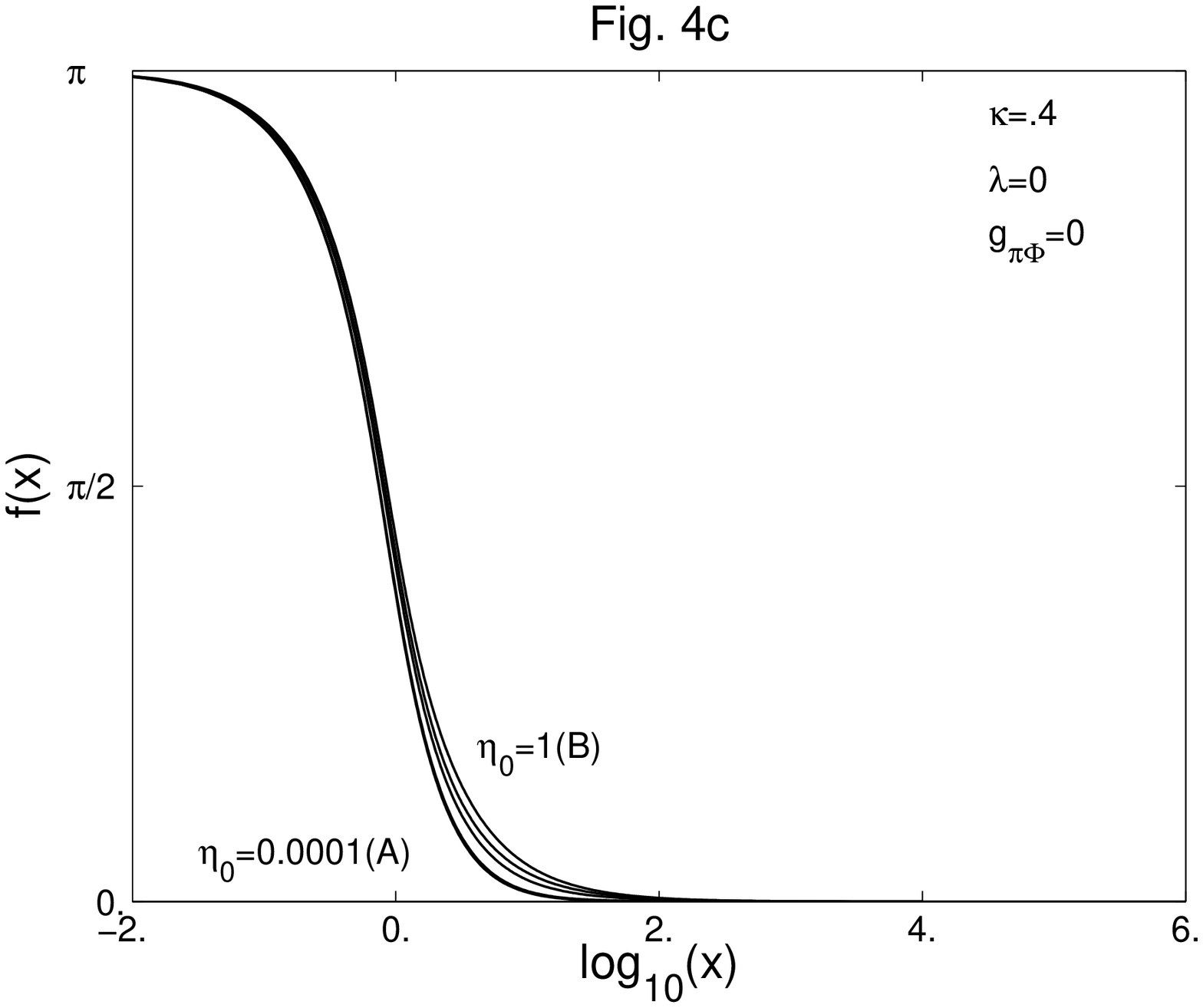}}
\caption{
(a) The gauge field function $a(x)$ is shown for several values of $\eta_0$ for 
fixed $\kappa=0.4$, $\lambda=0$ and $g_{\pi\Phi}=0$.
(b)
The same as Fig.~4a for the scaled Higgs field function $h(x)/\eta_0$.
(c)
The same as Fig.~4a for the chiral function $f(x)$. 
The different lines correspond to 
$\eta_0=1(B)$, $\eta_0=0.03(B)$, $\eta_0=0.03(\tilde A)$ , $\eta_0=0.03(A)$
and $\eta_0=0.0001(A)$ from top to bottom.}
\end{figure}

In Fig.~4a the profile of the gauge field function $a(x)$ is shown for 
$\eta_0=1.0$ and $\eta_0=0.03$ on branch $B$, for
$\eta_0=0.03$ on branch $\tilde A$, and, for $\eta_0=0.03$ and $\eta_0=0.0001$
on branch $A$. While for $\eta_0=1.0$, $\eta_0=0.03$ ($B$), 
and $\eta_0=0.03$ ($\tilde{A}$), $a(x)$ is a monotonically 
decreasing function of $x$, it develops a 
local maximum at some stage on the branch $\tilde{A}$ while $\eta_0$ increases. Passing
to the branch $A$, now with decreasing $\eta_0$, this local maximum persists. 
Along this path the local maximum $a(x_{\rm max})$ and its location $x_{\rm max}$
increase and reach $a(x_{\rm max}=\infty)=1$ as $\eta_0$ tends to zero, 
while the asymptotic region, where $a(x)$ decays to zero, is shifted to infinity.
Thus, in the limit $\eta_0 \rightarrow 0$ the gauge field function 
of the Monopole-Skyrmion solution tends to the corresponding function 
of the gauged Skyrmion solution for all $x$ {\it except} at infinity. Therefore the
convergence of the Monopole-Skyrmion on $A$ and $\tilde A$ to the gauged-Skyrmion
solutions on $A_{\rm gS}$ and $\tilde A_{\rm gS}$ is {\it non-uniform}.

In Fig.~4b we exhibit the profiles of the scaled Higgs function $h(x)/\eta_0$ 
for the same parameters $\eta_0$ like in Fig.~4a. We observe that $h(x)/\eta_0$ is a 
monotonically increasing function of $x$ for all $\eta_0$.
However along the path of values of $\eta_0$ described above, the magnitudes of the
functions $h(x)/\eta_0$ become progressively smaller on an increasing interval,
while the asymptotic region, where $h(x)$ approaches its vacuum value $\eta_0$,
is shifted to increasing values of $x$. 
In the limit $\eta_0 \rightarrow 0$ the function $h(x)$ vanishes everywhere,
signaling the merging of the Monopole-Skyrmions to the (Higgs-free) gauged-Skyrmions
in this limit.

In Fig.~4c we show the profiles of the chiral function $f(x)$ for the same values of
parameters $\eta_0$ as in Figs.~4a,b.
For all values of $\eta_0$ the function $f(x)$ in a monotonically decreasing
function of $x$.
In contrast to the gauge field function $a(x)$ and the Higgs field function
$h(x)$ the chiral function $f(x)$ does not change considerably with $\eta_0$. 

The case discussed above in Figs.~4a,b,c pertains to $  \kappa <  \kappa_A^{cr}$.
For $ {\kappa}_{\rm B}^{cr} < {\kappa}<  {\kappa}_{\rm A}^{cr} $
there are gauged-Skyrmion solutions of both types, namely those on 
branches $A_{\rm gS}$
and $\tilde A_{\rm gS}$ as well as on branch $B_{\rm gS}$, so
the convergence of the Monopole-Skyrmion solutions to the gauged-Skyrmion
solutions can be both uniform and non-uniform . In that case the Monopole-Skyrmion
solutions on the branches 
$A$ and $\tilde{A}$ approach the gauged-Skyrmion solutions on the 
corresponding branches $A_{\rm gS}$ and $\tilde{A}_{\rm gS}$ in a 
{\it non-uniform} way, for the same reasons as described above.
However, the  gauge field function of the Monopole-Skyrmion
solutions of the branch $B$ obey the
same asymptotic behaviour as the corresponding function of the 
gauged Skyrmion solutions of the branch $B_{\rm gS}$. For these solutions
the convergence is {\it uniform}.

For $  \kappa >   \kappa_A^{cr}$, there is only one type of gauged-Skyrmion
solution, namely those on branch $B_{\rm gS}$ for which the asymptotic value of the gauge
field function $a(x)$ equals zero like for the Monopole-Skyrmion solution, and hence the
convergence of these solutions as $\eta_0 \to 0$ is {\it uniform}.

\subsection{$\eta \gg f_\pi$}

Let us now consider the case where the scale $\eta$ is much larger than
the scale $f_\pi$. In Fig. 5 we show the field configurations for 
$\eta_0=2700$ (solid lines) and for $\eta_0=1$ (dashed lines) for comparison. 
We observe that the gauge field
function $a(x)$ for $\eta_0=2700$ approaches its asymptotic value at a very
small distance from the origin. The same applies to the scaled Higgs field
function $h(x)/\eta_0$, except for the long ranged tail, which is due to 
the power law decay for vanishing Higgs mass, i.~e. for $ {\lambda}=0$.
The chiral function $f(x)$, however, extends to larger distances from the origin.
This is in contrast to the configuration for $\eta_0=1$, 
where the change in the profile of all functions is roughly on the same 
interval. 
 
\begin{figure}[!h]
\centering
\mbox{\epsfxsize=10.cm\epsffile{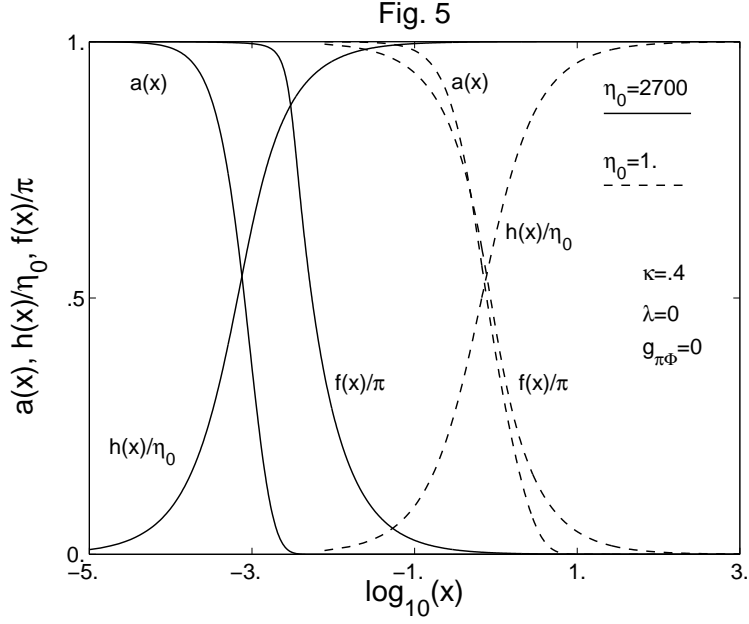}}
\caption{
The profiles of the gauge field function $a(x)$, the scaled Higgs field 
function $h(x)/\eta_0$ and the scaled chiral function $f(x)/\pi$ are 
shown for $\eta_0=2700$ (solid lines) and $\eta_0=1$ (dashed lines) for
fixed $\kappa=0.4$, $\lambda=0$ and $g_{\pi\Phi}=0$.}
\end{figure}

Note, that $\eta_0=2700$ corresponds to the case where the parameters 
$f_\pi$ and $\eta$ are of the magnitude of the pion decay constant in 
low energy QCD and 
the vacuum expectation value of the Higgs field in the Weinberg-Salam model, 
respectively. For the energy of this 
solution we find $4 \pi  {E}= 2700.013 \ 4 \pi$, i.~e. roughly the energy of the 
BPS Monopole ($4 \pi \eta $). 
An appealing physical interpretation of this solution seems to 
be that a Monopole resides at the center of a baryon and dominates its mass.

This can also be seen in a different way using gauge invariant quantities like
topological charges.
We define the topological Monopole charge density as
\begin{equation}
\tilde{\rho}_{MP} = \frac{1}{4\pi \eta} Tr\left\{ F_{ij}D_k \Phi \right\}
\varepsilon^{ijk} \ ,
\label{top_MP}
\end{equation}
and according to Ref.~\cite{AT,BT} the gauge invariant baryonic charge density as 
\begin{equation}
\tilde{\rho}_{B} = \frac{1}{12\pi^2}\left(
D_i\xi^a D_j\xi^b D_k\xi^c \xi^d \epsilon^{abcd} 
- 3 \xi^4 F^\alpha_{ij} D_k\xi^\alpha\right) \varepsilon^{ijk} \ , 
\label{top_b}
\end{equation}
where we have defined for $U=\exp(i \pi^\alpha \tau^\alpha)$
\begin{equation}
\xi^4 = \cos |\pi| \ , \ \xi^\alpha = \sin |\pi| \frac{\pi^\alpha}{|\pi|} \ ,
\ D_i\xi^\alpha= \partial_i \xi^\alpha  
+\varepsilon^{\alpha\beta\gamma} A_i^\beta \xi^\gamma  \ , \ 
\ D_i\xi^4= \partial_i \xi^4  \ , 
\nonumber 
\end{equation}
with $a, b, c, d$ run form $1$ to $4$ and  $\alpha, \beta, \gamma$ 
run from $1$ to $3$.
For the spherically symmetric Ansatz (\ref{A_an}-\ref{U_an}) and with the 
dimensionless coordinate $x$ the scaled charge densities become
\begin{equation}
\tilde{\rho}_{MP} = \frac{\left[h(1-a^2)\right]'}{4\pi x^2\eta_0} \ \ \ {\rm and} \ \ \
\tilde{\rho}_{B}=  -\frac{\left[f+(1-2 a^2)\sff\cff\right]'}{4\pi^2 x^2} \ .
\nonumber
\end{equation}
For $\eta_0 = 0.0001$ and  $\eta_0 =2700$,  
with fixed $ {\kappa}=0.4$, ${\lambda}=0$, ${g}_{\pi\Phi}=0$,
we show in Fig.~6 
the functions $\rho_{MP}=4\pi x^2 \tilde{\rho}_{MP}$ 
(solid lines) 
and $\rho_{B}=4\pi x^2 \tilde{\rho}_{B}$ (dashed lines) 
normalized by their respective maxima. 
\begin{figure}[!h]
\centering
\mbox{\epsfxsize=10.cm\epsffile{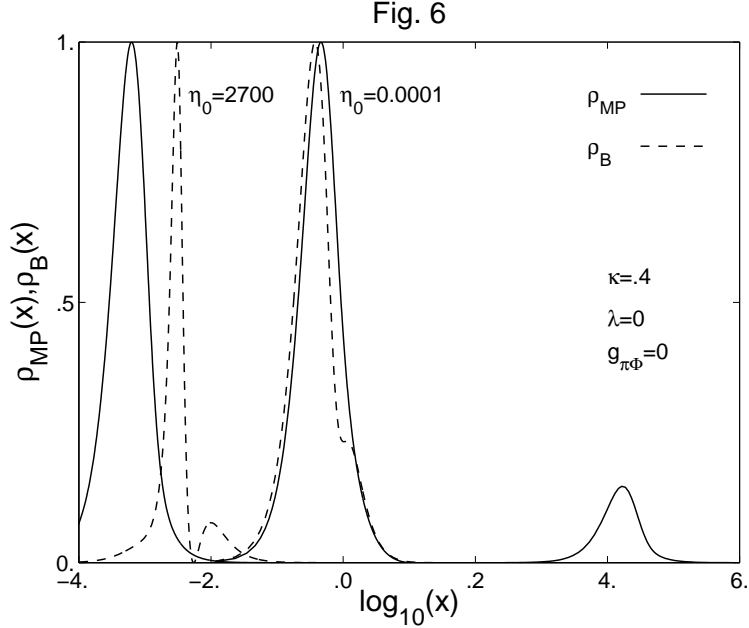}}
\caption{
The profiles of the normalized topological functions 
$\rho_{MP}(x)/{\rm max}(\rho_{MP})$
(solid lines) and 
$\rho_{B}(x)/{\rm max}(\rho_{B})$ (dashed lines) are shown for $\eta_0=2700$
and $\eta_0=0.0001$ for fixed $\kappa=0.4$, $\lambda=0$ and $g_{\pi\Phi}=0$.}
\end{figure}
The values of the normalization constants are 
$(4\pi x^2 \tilde{\rho}_{MP})_{\rm max} = 2.113$ 
and 
$(4\pi x^2 \tilde{\rho}_{B})_{\rm max} = 11080$
for $\eta_0=0.0001$,
and
$(4\pi x^2 \tilde{\rho}_{MP})_{\rm max} = 0.31$ 
and 
$(4\pi x^2 \tilde{\rho}_{B})_{\rm max} =0.113$ 
for $\eta_0=2700$, respectively.

>From Fig.~6 we observe that for $\eta_0=2700$ the location 
$x^{\rm MP}_{\rm max}$ of the maximum 
of the function $\rho_{MP}$ resides at a considerably smaller distance 
from the origin than the 
the location $x^{\rm B}_{\rm max}$ of the maximum of the function
$\rho_{B}$. This confirms the interpretation as a Monopole inside 
a baryon. Note, that the function $\rho_{B}$ possesses an additional
local maximum at a larger distance from the origin than the global maximum
and a minimum with vanishing magnitude between the both maxima. 
We found that half the baryonic charge stems from the area below the 
first peak and half from the area below the second peak. 
This can be understood as follows.
For distances larger than or near the location of the minimum $x_{\rm min}$ 
the gauge field function $a(x)$ almost vanishes. Setting $a(x)=0$ in 
$\rho_{B}$ we find that the minimum corresponds to 
$f(x_{\rm min})=\pi/2$.
Splitting the integral of the baryonic charge density into two parts,
\begin{equation}
B = \int_0^{x_{\rm min}} \rho_{B} dx 
+ \int_{x_{\rm min}}^\infty \rho_{B} dx
\nonumber
\end{equation}
we find for both parts the value $1/2$.

Let us next discuss the charge densities for $\eta_0=0.0001$.
In this case we observe from Fig.~6 that the peaks of $\rho_{MP}$ and
$\rho_{B}$ are roughly at the same location. However, at a large distance from 
these peaks the function $\rho_{MP}$ possesses a second local maximum.
We found, that the magnetic charge stems mainly from this second local maximum,
whereas the contribution from the global maximum is marginal.
This behaviour becomes plausible if we consider the profile of the 
gauge field function $a(x)$ in Fig.~4a. For small values of $x$ 
the function $a(x)$ is close to one. Consequently, the magnetic charge density 
is small. For larger values of $x$ the function $a(x)$ develops a local 
minimum, this leads to the first peak of $\rho_{MP}$. 
When $a(x)$ approaches again the value one, the function $\rho_{MP}$ 
again becomes very small. At large distances the function $a(x)$ decays 
to its asymptotic value. This leads to the second maximum of $\rho_{MP}$.
However, the magnetic charge density also depends on the Higgs field function 
$h(x)$, and one would expect that the charge density has to be small if 
$h(x)$ is small, i.~e. in the region where the gauge field function possesses 
its local minimum, see Figs.~4a and 4b.
This is indeed the case. Note, that for $\eta_0=0.0001$ the normalization
constants of the function $\rho_{B}$ is several orders of magnitude
larger then the normalization constant of the function $\rho_{MP}$.
Thus, the Monopole charge density at the first peak is indeed very small 
compared to the baryon charge density.

In analogy to the interpretation of a Monopole inside a baryon for 
large values of $\eta_0$, one could interpret the solutions for 
small values of $\eta_0$ as a baryon inside a Monopole. However, the 
gauge field is nontrivial at the location of the baryon and the Higgs field
is not in the vacuum in this region. Thus the baryon would reside in a nearly
symmetric phase. One may speculate, that this scenario might be interesting
in respect to the decay of the baryon.

\subsection{ $g_{\pi\Phi}>0$ }

In most calculations we fixed the parameter $g_{\pi\Phi}=0$. In view of the 
discussion in section 2 this needs some clarification.
The interaction term of the chiral matrix with the Higgs field was introduced
into the model basically for ideological reasons as it fixes
the value of chiral function at infinity,  $f_\infty$ = 0 (say).
Then we assumed that in the limit $g_{\pi\Phi}\rightarrow 0$ the 
asymptotic value of the chiral function is stil fixed,
and that the field configurations behave smoothly.
Indeed, we found from our numerical analysis that this is the case 
and that the assumption is justified.

Let us now discuss the case where $g_{\pi\Phi}$ is finite.
As long as $g_{\pi\Phi}$ is small,
the dependence of the solutions on the parameters $\eta_0$ and $\kappa$ 
does not change considerably.
In particular the bifurcation pattern for small values of 
$\eta_0$ as shown in Figs.~3 persists for small $g_{\pi\Phi}$.
The reason is simply that  $g_{\pi\Phi}$ enters the differential equations
only as a factor of the Higgs field function $h(x)$. In the limit 
$\eta_0 \rightarrow 0$ the Higgs field function vanishes and consequently
the equations do not depend on $g_{\pi\Phi}$ in this limit.
To discuss the more general case of finite $\eta_0$ let us assume 
that for some parameters $\kappa,\lambda,\eta_0$ solutions on the branches $A$, $B$ and
$\tilde{A}$ coexist and form a butterfly in the 
$E-\eta_0$ diagram for $g_{\pi\Phi}=0$. Then the 
butterfly will persist for small values of $g_{\pi\Phi}$.
As $g_{\pi\Phi}$ increases, the butterfly shrinks in size and disappears
at a critical value of $g_{\pi\Phi}$, e.~g. $g^{cr}_{\pi\Phi}\approx 0.4$
for fixed $\kappa=0.4$ and $\lambda=0$.
Thus, the bifurcation pattern is similar to Fig.~3, if we replace
$\kappa$ by $\eta_0$, $\eta_0$ by $g_{\pi\Phi}$ and interchange the
role of $A$ and $B$.

We now consider the case where $g_{\pi\Phi}$ becomes very large
at fixed parameters $\kappa,\lambda,\eta_0$. 
In the limit $g_{\pi\Phi}\rightarrow \infty$
the potential (\ref{Vpot}) becomes a constraint, which for the spherically
symmetric Ansatz (\ref{Phi_an}), (\ref{U_an}) becomes 
\begin{equation}
h^2(x)\sqff(x) \longrightarrow 0
\ \ {\rm as } \ \ g_{\pi\Phi}\rightarrow \infty \ .
\end{equation}
Note, that this constraint can neither be solved by a vanishing Higgs 
field function, $h(x) \equiv 0$, because this violates the boundary condition 
$h(x\rightarrow\infty )= \eta_0$, nor by a constant chiral function, 
$f(x)\equiv 0$ or $\pi$, because this violates boundary conditions at 
the origin or at infinity. However, there is a third possibility.
If the Higgs field function vanishes on the interval $[0,x_0]$ and 
the chiral function vanishes on the interval $[x_0,\infty]$ then 
the function $h^2(x)\sqff(x)$ vanishes everywhere. 

\begin{figure}[!h]
\centering
\mbox{\epsfxsize=10.cm\epsffile{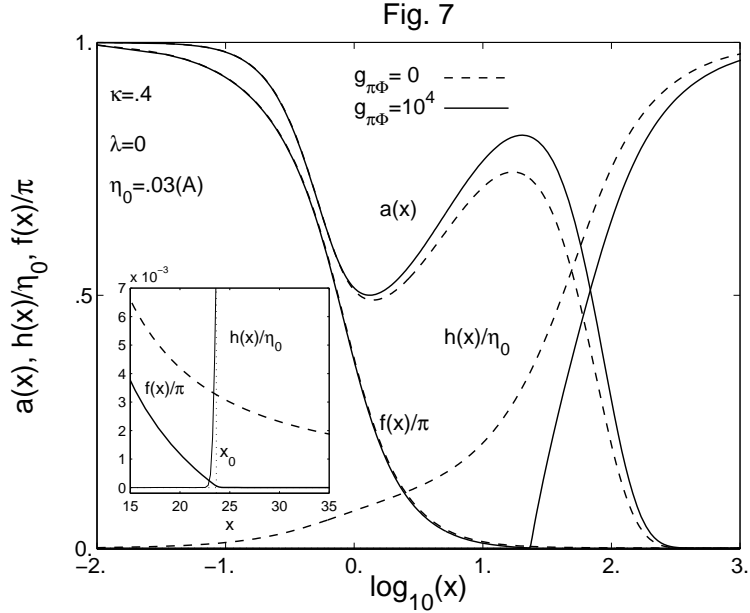}}
\caption{
The profiles of the gauge field function $a(x)$, the scaled Higgs field 
function $h(x)/\eta_0$ and the scaled chiral function $f(x)/\pi$ are 
shown for $g_{\pi\Phi}=10^4$ (solid lines) and $g_{\pi\Phi}=0$ (dashed lines) 
for fixed $\kappa=0.4$, $\lambda=0$ and $\eta_0=0.03$ an the branch $A$.}
\end{figure}

In Fig.~7 we 
show the field configurations for $g_{\pi\Phi}=10^4$ and for $g_{\pi\Phi}=0$
for comparison for fixed parameters $\kappa=0.4$, $\lambda=0$ and 
$\eta_0=0.03$ on the 
branch $A$. We observe that the Higgs field function $h(x)$ is indeed 
almost zero on the interval $[0,x_0]$, with $x_0\approx 23.65$, while 
the chiral function $f(x)$  changes continuously from $f(0)=\pi$ 
to $f(x_0)\approx 0$.
On the interval $[x_0,\infty]$ the chiral function is almost zero and 
the Higgs field function changes continuously form  $h(x_0)\approx 0$  to 
its asymptotic value $h(\infty)=\eta_0$. 
The figure suggests that
in the limit $g_{\pi\Phi} \rightarrow \infty$ the derivatives of 
the functions $h(x)$ and $f(x)$ will be finite but non-continuous at $x_0$,
whereas the gauge field function remains twice differentiable at $x_0$.

Taking into account the behaviour of the functions for large $g_{\pi\Phi}$,
we observe
from the differential equation for the chiral function $f(x)$ (\ref{deqs0}),
that the $\pi\Phi$ interaction term will be almost zero for $x<x_0$. Assuming
that $h(x)$ increases linearly at $x_0$, we find for large $g_{\pi\Phi}$ 
that for $x>x_0$ the chiral function decays exponentially with the 
exponent $\sim -g_{\pi\Phi}$. Hence, we find the following scenario.
For $x<x_0$ the interaction of the Higgs field with the chiral field vanishes,
whereas for $x>x_0$ the chiral field becomes increasingly massive. 
Consequently, the baryon is trapped inside the Monopole.
On the other hand, because the magnetic charge density is proportional
to the Higgs field, there will be (almost) vanishing Monopole density for 
$x<x_0$ for large $g_{\pi\Phi}$ and the Monopole is expelled from the baryon.

\section{Normal modes}

To show the instability of a solution of Eqs.~(\ref{deqs0}) 
we determine the eigenvalues of the fluctuation matrix around that solution. 
The existence of a negative eigenvalue of a normalizable fluctuation mode
indicates that a deformation of the solutions in the direction of this
mode lowers its energy. Hence this solution can not be stable.
Therefore, to show the instability of a solution, it is sufficient to
find a normalizable fluctuation mode with negative eigenvalue.
For the discussion of the normal modes we will adopt the
methods discussed in Refs.~\cite{AKY2,BGKK}.

Here we consider only radial fluctuations around the solutions.
We use the dimensionless coordinate $x$ from the beginning.
We introduce into the Ansatz small space-time dependent fluctuations 
$\Psi_\alpha(x)e^{i\omega t}$, $\alpha=a,b,c,h,f$,
\begin{eqnarray}
a(x) & \rightarrow & a(x) + \Psi_a(x)e^{i\omega t} \ ,\nonumber\\
b(x) & \rightarrow & b(x) + \Psi_b(x)e^{i\omega t} \ , \nonumber\\
c(x) & \rightarrow & c(x) + \sqrt{2}\Psi_c(x)e^{i\omega t} \ , \nonumber\\
h(x) & \rightarrow & h(x) + \frac{\Psi_h(x)}{x}e^{i\omega t}  \ ,\nonumber\\
f(x) & \rightarrow & f(x) + \frac{\Psi_f(x)}{x}e^{i\omega t}  \ ,
\label{fluc}
\end{eqnarray}
and expand the Hamiltonian and Lagrangian to second order in the 
fluctuation function $\Psi_\alpha$.
As we are interested in the fluctuations around the static solutions, 
we set $b \equiv 0$, $c \equiv 0$ and extremize
the Lagrangian in the background of the functions $a,h,f$, i.~e.
whenever second derivatives of these functions appear, they are replaced 
by the right hand side of the differential equations (\ref{deqs0}).

The form invariance of the Ansatz (\ref{A_an})-(\ref{U_an}), expressed by 
Eqs.~(\ref{invarg}), (\ref{invarf}), reflects itself
in the existence of normal modes with vanishing energy eigenvalue. 
These gauge zero modes obey the conditions
\begin{equation}
\Psi_a=0 \ , \ \Psi_c= \frac{x}{\sqrt{2}}\left(\frac{\Psi_b}{a}\right)' \ , \ 
\Psi_h=0 \ , \ \Psi_f=0 \ .
\label{gz_modes}
\end{equation}
Because we are interested in the non-zero modes, we want to exclude 
the gauge zero modes. This can be done by imposing the conditions 
that the normal modes have to be orthogonal to the gauge zero modes with 
respect to the metric 
\begin{equation}
<\tilde{\Psi},\Psi>_{bc}=
      \int{\left(\tilde{\Psi}_b \Psi_b + \tilde{\Psi}_c\Psi_c\right)
                    + \dots } \ \ dx \ .
\label{metric_bc}
\end{equation}
(We will give the complete form of the metric later).
This leads to the condition on the functions $\Psi_b,\Psi_c$ 
\begin{equation}
K[\Psi_b,\Psi_c]=\left(x \Psi_c\right)'-\sqrt{2}a \Psi_b = 0  \ .
\label{orth_cond}
\end{equation}
To exclude the gauge zero modes we add $\mu K^2$ to the 
Lagrangian, where $\mu$ is a Lagrange multiplier.

>From the system of differential equations we find, that 
the functions $\Psi_b$ and $\Psi_c$ couple to each other, but 
not to the functions $\Psi_a$, $\Psi_h$ and $\Psi_f$ and vice versa.
Thus, we have two decoupled systems of differential equations, which can
be solved separately.

\subsection{The system $\{\Psi_b \Psi_c\}$}

Let us first address the system $\{\Psi_b \Psi_c\}$. 
The differential equations become
\begin{eqnarray}
\Psi_c''+\omega^2 \Psi_c & = &  
2 \frac{1+a^2}{x^2} \Psi_c + 2\sqrt{2}\frac{x a' -a}{x^2}\Psi_b \ ,
\label{sys_bc1} \\
\Psi_b''+\omega^2 \Psi_b  & = & 
   V_{bb}\Psi_b + 2\sqrt{2}\frac{x a'-a}{x^2}\Psi_c \ ,
\label{sys_bc2}
\end{eqnarray}
and the corresponding static Hamiltonian is 
\begin{eqnarray}
{\cal H}_{bc}[\Psi_b, \Psi_c] & = &  
\frac{4\pi f_\pi}{g}
\int \left\{V_{bb} \Psi_b^2 + \Psi_b'^2 +\frac{1+2a^2}{x^2}\Psi_c^2+ \Psi_c'^2
+2 \frac{\Psi_c'\Psi_c}{x}
\right.
\nonumber\\
& &
\left.
+2\sqrt{2}\left[\left(\frac{a}{x}\right)' \Psi_b \Psi_c 
                    -\frac{a}{x}(\Psi_b \Psi_c)'\right]
\right\}dx\label{Hbc}  
\end{eqnarray}
with
\begin{equation}
V_{bb} = 
             \left(\frac{3 a^2-1}{x^2}+ h^2 +\sqff +
             4 {\kappa}^2\sqff \left(f'^2 + \frac{a^2\sqff}{x^2}
             \right)  \right) \ ,
\end{equation}
where we have  set the Lagrange multiplier $\mu$ equal to $\frac{1}{2}$.
The boundary conditions for the functions $\Psi_c$ and $\Psi_b$
are given by
\begin{eqnarray}
x=0 & : & \Psi_c=0 \ , \ \Psi_b=0 \ , 
\nonumber\\
x\rightarrow \infty & : & \Psi_c\rightarrow 0 \ , \ \Psi_b\rightarrow 0  \ . 
\label{BC_bc}
\end{eqnarray}

For solutions of (\ref{sys_bc1}), (\ref{sys_bc2}) we can evaluate the 
energy integral (\ref{Hbc}) by integration by parts and using
(\ref{sys_bc1}), (\ref{sys_bc2}) and (\ref{BC_bc}).
We find for the dimensionless energy 
$E_{bc}= \frac{g}{4\pi f_\pi}{\cal H}_{bc}$,
\begin{equation}
E_{bc}= \omega^2 \ ,
\nonumber
\end{equation}
if we assume that the functions are normalised with respect to the 
metric (\ref{metric_bc}). Thus, $\omega^2$ denotes the energy eigenvalue in
units of $4\pi f_\pi/g$.

We solved the system (\ref{sys_bc1}), (\ref{sys_bc2}) for several values of
the parameters $\eta_0$, $\kappa$, $\lambda$ and $g_{\pi\Phi}$ and found 
only solutions with $\omega^2$ positive.
For fixed values of the parameters we found several discrete normal modes
which can be characterized by the number of nodes $N$ of the fluctuation
function $\Psi_b$. Their eigenvalues $\omega_N^2$ increase with the number of
nodes $N$. The lowest positive eigenvalue $\omega_1^2$ corresponds
to one node of the function  $\Psi_b$, see Fig.~8 (inlet). 
It seems to be likely that the system $\{\Psi_b\Psi_c\}$ 
possesses an infinite number of discrete positive eigenvalues, 
forming a sequence with convergence to $\eta_0^2$. 

\begin{table}[!h]
\begin{center}
\begin{tabular}{|c|cc|cc|} 
 \hline 
\multicolumn{1}{|c|} { $ $ }& 
\multicolumn{2}{|c|} { $\eta_0=0.032$ ($\tilde{A}$) } &
\multicolumn{2}{|c|} { $\eta_0=0.026$ ($\tilde{A}$) } \\
 \hline 
$N$ & $\omega_N^2$ ($\times 10^{-3}$) & fit 
    & $\omega_N^2$ ($\times 10^{-3}$) & fit \\
 \hline 
1  & 0.9767 & 0.9608 & 0.66033 & 0.65785 \\
2  & 1.0098 & 1.0080 & 0.67165 & 0.67144 \\
3  & 1.0173 & 1.0169 & 0.67400 & 0.67400 \\
4  & 1.0201 & 1.0200 & 0.67486 & 0.67486 \\
5  & 1.0215 & 1.0214 & 0.67527 & 0.67527 \\
6  & 1.0222 & 1.0222 & 0.67549 & 0.67549 \\
7  & 1.0227 & 1.0227 & 0.67563 & 0.67563 \\
8  & 1.0230 & 1.0230 & 0.67571 & 0.67571 \\
 \hline 
\end{tabular}
\end{center} 
{\bf Table 1}\\
The eight lowest positive eigenvalues for $\eta_0=0.032$ and $\eta_0=0.026$
on the sphaleron branch $\tilde{A}$ for $\kappa=0.4$, $\lambda=0$ and 
$g_{\pi\Phi}=0$ together with the fitted eigenvalues.
\end{table}

For large $N$ the eigenvalues $\omega_N^2$ can be well approximated 
by the formula
\begin{equation}
\omega_N^2 = \left(\eta_0^2 - \frac{C}{N^2}\right)^2 \ ,
\label{om_fit}
\end{equation}
where $C$ depends on the parameters $\eta_0$, $\kappa$, $\lambda$ and
$g_{\pi\Phi}$.
In Table 1 we give the first eight eigenvalues together with their fitted
values for $\eta_0=0.032$ and $\eta_0=0.026$ on the branch $\tilde{A}$ 
for fixed $\kappa=0.4$, $\lambda=0$ and $g_{\pi\Phi}=0$. For the 
constants $C$ we found $C(\eta_0=0.032)=0.001$ and 
$C(\eta_0=0.026)=0.00035$.

In Fig.~8 we show the lowest positive eigenvalue $\omega_1^2$ as a function
of $\eta_0$ for $ {\kappa}=0.4$, $ {\lambda}=0$ and $ {g}_{\pi\Phi}=0$.
For all branches $B$, $A$ and $\tilde{A}$ 
the eigenvalue is a monotonically 
increasing function of $\eta_0$ and vanishes in the limit 
$\eta_0 \rightarrow 0$.
 
\begin{figure}[!h]
\centering
\mbox{\epsfxsize=10.cm\epsffile{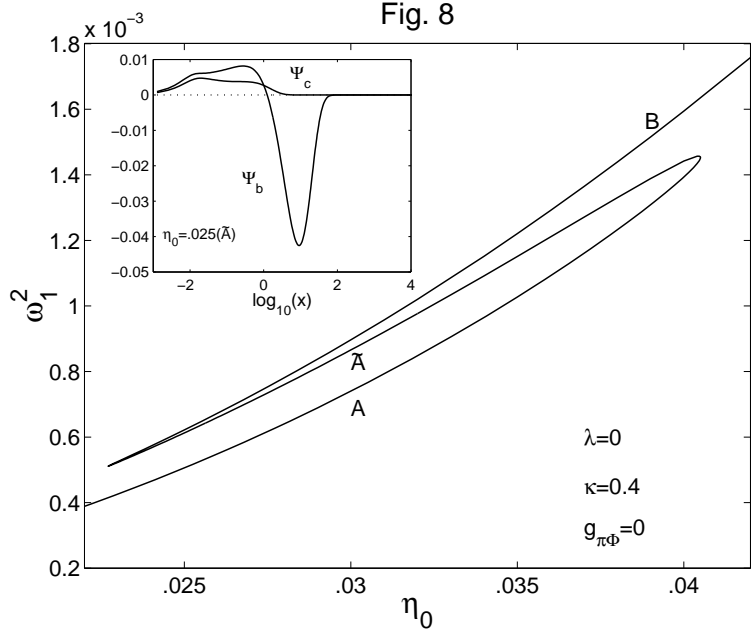}}
\caption{
The lowest positive eigenvalue $\omega_1^2$ is shown as a function of
$\eta_0$ for fixed $\kappa=0.4$, $\lambda=0$ and $g_{\pi\Phi}=0$.
The inlet shows the fluctuation functions $\Psi_b(x)$ and $\Psi_c(x)$ 
for $\eta_0=0.025$ on the branch $\tilde{A}$.}
\end{figure}

\subsection{The system $\{\Psi_a\Psi_h\Psi_f\}$}

For the system $\{\Psi_a\Psi_h\Psi_f\}$ the gauge zero modes are absent 
and we can calculate the differential equations and the static Hamiltonian 
directly. The system of differential equations is given by
\begin{eqnarray}
\Psi_a'' +\omega^2 \Pa & = &
\left[ 
\frac{3a^2-1}{x^2} +h^2 + \sqff 
+4 \hk\sqff(f'^2+\frac{3 a^2\sqff}{x^2})
\right]\Pa
\nonumber\\
& & 
+\left[\frac{2ah}{x}\right]\Ph
+\left[\frac{8\hk a f' \sqff}{x}\right] \Pf'
\nonumber\\
& & 
+\frac{2}{x}\left[a \cff\sff +4\hk a \sff \left(\cff f'^2 - \frac{\sff f'}{x}
+\frac{2 a^2\cff \sqff}{x^2}\right)\right] \Pf
\\
\Ph'' +\omega^2 \Ph & = &
\left[
\frac{2a^2}{x^2}+ {g}_{\pi\Phi}^2\sff + 4\hk(3 h^2-\eta_0^2) 
\right]\Ph
+\left[\frac{4 a h}{x}\right]\Pa + \left[2 {g}_{\pi\Phi}^2 \cff\sff h\right] \Pf
\\
\Pf''+\omega^2 \Pf & = &
\left[
\frac{4}{x G}
\left(
a \cff \sff x^4
+32 {\kappa}^4 a^2 \sqqff(a^3 \cff\sff +a'f'x^2)
\right.  \right. 
\nonumber \\
& &
\left. \left.  
+4\hk\sff x^2\left(
2 a^3 \cff \sqff - a \cff f'^2 x^2 +2 a \sff f' x- a' \sff f' x^2 
\right.  \right. \right.
\nonumber \\
& &
\left. \left.  \left.
- {g}_{\pi\Phi}^2 a \cff \sqff h^2 x^2\right)
\right)
\right] \Pa 
\nonumber \\
& &
-\left[16 \hk \frac{a \sqff f' x}{8\hk a^2\sqff+x^2}\right] \Pa'
+\left[2  {g}_{\pi\Phi}^2 \frac{\cff \sff h x^2}{8\hk a^2\sqff+x^2}\right] \Ph
\nonumber \\
& &
+
\left[\frac{1}{x^2 G}
\left(
x^4(2 a^2+  {g}_{\pi\Phi}^2 h^2 x^2 )(1 - 2 \sqff) 
\right.  \right.
\nonumber \\
& &
\left. \left.
+
64  {\kappa}^4 a^3 \sqff \left\{ a \sqff ( a^2 (1 - 2 \sqff)-2)
+ 2 a \cff \sff f'x 
\right.  \right. \right.
\nonumber \\
& &
\left. \left. \left.
+ a f'^2 x^2 + 2 a' x \sqff \right\}
\right. \right.
\nonumber \\
& &
\left. \left.
+8 \hk a x^2 \left\{
a \sqff ( a^2 (1-4 \sqff )-2 )
+6 a \cff\sff f' x + 2 a \sqff x^2 f'^2 
\right. \right. \right.
\nonumber \\
& &
\left. \left. \left.
-a x^2 f'^2-4 \cff \sff a'f'x^2
+2 \sqff a'x - {g}_{\pi\Phi}^2 a \sqff h^2 x^2 \right\}
\right) 
\right] \Pf
\nonumber \\
& &
+
\left[
\frac{16 \hk }{x G}
a \sff \left((x^2 + 8 \hk a^2 \sqff) ( a(\sff-\cff f' x) -\sff a'x)
\right)
\right] \Pf' 
\end{eqnarray}
with 
\begin{equation}
G= x^4 +16\hk a^2\sff^2(x^2+4\hk a^2\sff^2 ) \ . 
\end{equation}

We now find for the dimensionless energy of
the solutions of the differential equations
\begin{equation} 
E_{ahf} =  \omega^2 \int\left\{\Psi_a^2 +\frac{\Psi_h^2}{2} 
                         +\left[1+\frac{8 {\kappa}^2 a^2 \sqff}{x^2}\right]
                            \frac{\Psi_f^2}{2} \right\} dx \ .
\nonumber 
\end{equation} 
>From this form we can define an appropriate metric for the fluctuations $\Psi$ by
\begin{equation}
<\tilde{\Psi},\Psi>
=
\int\left\{\tilde{\Psi}_a\Psi_a + \tilde{\Psi}_b\Psi_b +\tilde{\Psi}_c\Psi_c
            + \frac{1}{2} \left(\tilde{\Psi}_h \Psi_h 
            +\left[1+\frac{8 {\kappa}^2 a^2 \sqff}{x^2}\right]
            \tilde{\Psi}_f\Psi_f\right)\right\} dx \ .
\label{metric}
\end{equation}
Assuming the normalisiation of the fluctuation functions according to 
(\ref{metric}), we find the energy of the solutions to be
$E_{ahf} =  \omega^2$,
i.~e. $\omega^2$ is again the energy eigenvalue.

\begin{figure}[!h]
\centering
\mbox{\epsfxsize=10.cm\epsffile{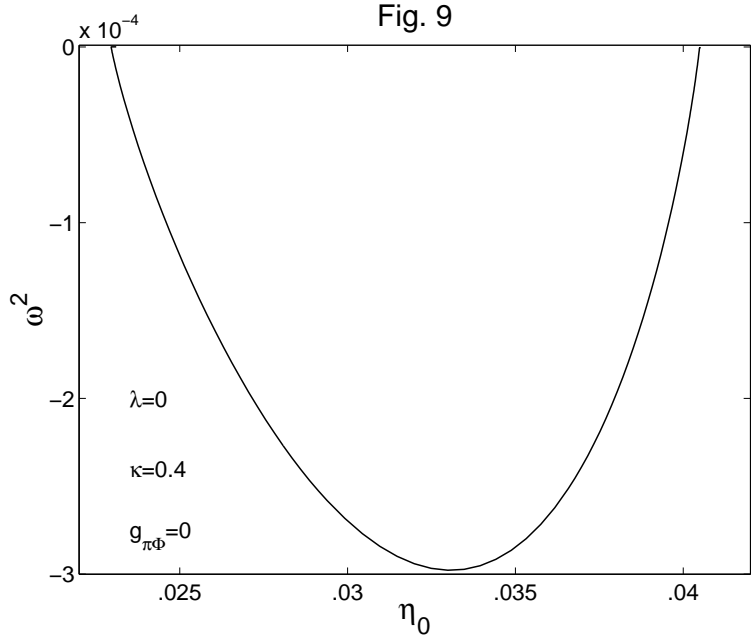}}
\caption{
The negative eigenvalue $\omega^2$ is shown as a function of $\eta_0$ 
for fixed $\kappa=0.4$, $\lambda=0$, $g_{\pi\Phi}=0$.}
\end{figure}

For the system $\{\Psi_a\Psi_h\Psi_f\}$ we found normalizable solutions 
only on the saddlepoint branch $\tilde{A}$. 
For these normal modes the eigenvalue $\omega^2$ is negative
and vanishes at the bifurcation points.
Hence these normal modes represent an instability mode of the saddlepoint
solutions.

In Fig.~9 we show the negative eigenvalue $\omega^2$ as a function
of $\eta_0$ for $ {\kappa}=0.4$, $ {\lambda}=0$ and $ {g}_{\pi\Phi}=0$.

\section{Summary and Discussion}

We have studied a model combining the Georgi-Glashow and the Skyrme
systems interacting mainly through the $so(3)$ gauge field, with an
additional interaction term between the Higgs and chiral fields. This
interaction term, which breaks the chiral symmetry and fixes the
asymptotic value of the chiral field at infinity, exploits the non
vanishing VEV of the Higgs field in an essential way.

The main emphasis of the study is the numerical analysis of the classical
solutions of the model, which may be relevant to the semiclassical
approach to monopole catalysis of Baryon decay~\cite{CW}.
The most interesting
feature of these solutions, both intrinsically and for the physical
reason just mentioned, is the particular bifurcation patterns that they
exhibit. These patterns are connected to similar bifurcations in the
solutions of the gauged Skyrme model, not involving a Higgs field, which
has led us to make a systematic study of the relation between the
solutions of these two models. This has been carried out by considering
particular limits, in terms of the independent parameters involved in
the Monopole-Skyrme model, in which the solutions of this model merge
with the solutions of the Higgs independent gauged Skyrme model.

The above mentioned technical investigations form the centre of gravity
of the present work. From the results obtained, some interesting
observations of physical relevance can be made.

One is the particular shape
of the bifurcations the solutions exhibit, namely what we have referred
to as 'butterfly' in the text. These are reminiscent of the kind of
bifurcations appearing in first order phase transitions. Unfortunately, in
the present form of the model considered, we have not been able to make
a concrete description for this phenomenon. This aspect of the work
is under consideration.

The other observation can be made more quantitatively. It was shown that
in some limit of the parameters, the Baryon resides in the core of the
monopole while in the other limit the converse, namely that the monopole
resides inside the Baryon. In particular, for the values of the
parameters fixed by the phenomenological values of the Higgs VEV and the
pion decay constants, exactly one half of the Baryon charge resides inside
the monopole core, and the rest outside.

\bigskip

{\bf \large Acknowledgements}
B.~K. was supported 
by Forbairt grant SC/97-636;
F.~Z. was supported HEA postdoctoral fellowship.
\small{

 }

\end{document}